\documentclass[aps,prc,showpacs,twocolumn,nofootinbib,groupedaddress,preprintnumbers]{revtex4-1}%preprint
\usepackage{color}
\usepackage{calligra}
\usepackage{longtable}
\usepackage{graphicx}% Include figure files
\usepackage{dcolumn}% Align table columns on decimal point
\usepackage{bm}% bold math
\usepackage{amsmath,amssymb}
\usepackage{mathrsfs}
\usepackage{color}
\usepackage{multirow}
\usepackage{subfigure}

\begin{document}

\preprint{KUNS-2638, YITP-16-100}


\title{Theoretical study of the $\Xi(1620)$ and $\Xi(1690)$ resonances in $\Xi_c \to \pi^+ MB$ decays}


\author{Kenta~Miyahara}
\email[]{miyahara.kenta.62r@st.kyoto-u.ac.jp}
\affiliation{Department of Physics, Graduate School of Science, Kyoto University, Kyoto 606-8502, Japan}
\author{Tetsuo~Hyodo}
%\email[]{}
\affiliation{Yukawa Institute for Theoretical Physics, Kyoto University, Kyoto 606-8502, Japan}
\author{Makoto~Oka}
%\email[]{}
\affiliation{Department of Physics, Tokyo Institute of Technology, Tokyo 152-8551, Japan\\
Advanced Science Research Center, Japan Atomic Energy Agency, Tokai, Ibaraki, 319-1195, Japan}
\author{Juan~Nieves}
%\email[]{}
\affiliation{IFIC, Centro Mixto Universidad de Valencia-CSIC Institutos de Investigaci\'on de Paterna, Aptdo. 22085, 46071 Valencia, Spain}
\author{Eulogio~Oset}
%\email[]{}
\affiliation{Departamento de F\'{\i}sica Te\'orica and IFIC, Centro Mixto Universidad de Valencia-CSIC,
Institutos de Investigaci\'on de Paterna, Aptdo. 22085, 46071 Valencia,
Spain}
%\thanks{}


\date{\today}

%%%%%%%%%%%%%%%%%%%%%%%%%%%%%%%%%%%%%%%%%%%%%%%%%%%%%%%%%%%%%%%%%%%%%%%%
\begin{abstract}       
Nonleptonic weak decays of $\Xi_c$ into $\pi^+$ and a meson $(M)$-baryon $(B)$ final state, $MB$, 
are analyzed from the viewpoint of probing $S=-2$ baryon resonances, i.e., $\Xi(1620)$ and $\Xi(1690)$, of which spin-parity and other properties are not well known.
We argue that the weak decay of $\Xi_c$ is dominated by a single quark-line diagram, preferred by 
the Cabibbo-Kobayashi-Maskawa coefficient, color recombination factor, the diquark correlation, 
and the kinematical condition.
The decay process has an advantage of being free from meson resonances in the $\pi^+M$ invariant mass distribution.
The invariant mass distribution of the meson-baryon final state is calculated with three different chiral unitary approaches,
assuming that the $\Xi(1620)$ and $\Xi(1690)$ resonances have $J^P=1/2^-$.
It is found that a clear peak for the $\Xi(1690)$ is seen in the $\pi\Xi$ and $\bar{K}\Lambda$ spectra. 
We also suggest that the ratios of the $\pi\Xi$, $\bar{K}\Lambda$ and $\bar{K}\Sigma$ final states 
are useful to distinguish whether the peak is originated from the $\Xi(1690)$ resonance or it is a $\bar{K}\Sigma$ threshold effect. 
                    
%We propose the $\Xi_c$-decay  emitting a $\pi^+$ and an additional meson-baryon pair ($MB$) to study  the 
%$\Xi(1620)$ and $\Xi(1690)$ resonances, which main properties are not well determined yet. 
%This decay process has the advantage of being free from meson 
%resonances in the $\pi^+M$ invariant mass distribution.  
%The analysis of this reaction shows there exists a dominant diagram,  considering  Cabibbo-Kobayashi-Maskawa matrix and color suppressions, 
%diquark correlations and  kinematical restrictions. The invariant mass distribution of the final meson-baryon is calculated 
%using three different  chiral unitary approaches, and found to present in all cases a clear peak   
%in the $\pi\Xi$ and $\bar{K}\Lambda$ decay channels for the $\Xi(1690)$. We suggest  that the ratios of the $MB$ decay fractions might
%be useful to distinguish whether the peak is originated from the $\Xi(1690)$ resonance or it is a $\bar{K}\Sigma$ threshold effect. 

\end{abstract}

  %	Kaon-baryon interactions,Baryons (including antiparticles) (for decays of baryons, see 13.30.-a,Chiral symmetries

% 11.55.Bq Analytic properties of S matrix
% 03.65.Nk Scattering theory

%\keywords{}

\maketitle

%=====================================================================

\section{Introduction}   \label{sec:intro}  %\ref{sec:intro}
The advent of the LHCb and its unexpectedly successful contribution to hadron physics has provided 
this area with a plethora of new reactions that have stirred a revival of hadron studies. Thus in LHCb, Belle, BESIII, and other facilities, 
new reactions and decays of heavy hadrons  have come under 
experimental study~\cite{Stone:2015iba}, which has triggered a large theoretical activity as well~\cite{Oset:2016lyh}. One of the 
interesting unexpected findings was the observation of two structures in the $J/\psi p$ invariant mass 
distribution in the $\Lambda_b \to J/\psi K^-p$ decay in Refs.~\cite{Aaij:2015tga,Aaij:2015fea} that were 
ascribed to two pentaquarks states. Prior to this experimental observation, the  $\Lambda_b \to J/\psi K^-p$ 
reaction was studied theoretically in Ref.~\cite{Roca:2015tea} and mass distributions associated to the 
production of the $\Lambda(1405)$ in the $K^- p $ and $\pi \Sigma$  spectra were predicted. 
In particular, the calculated $K^- p $ distribution  was in good agreement with the experimental findings. 
Furthermore, this information, together with predictions made for hidden charm states of 
$\bar{D}^*\Sigma_c-\bar{D}^*\Sigma_c^*$ molecular nature in Refs.~\cite{Wu:2010jy,Wu:2010vk, Xiao:2013yca} prompted a likely 
explanation in Ref.~\cite{Roca:2015dva} for the narrow state found in Refs.~\cite{Aaij:2015tga,Aaij:2015fea} (see 
also related works along the same line in Refs.~\cite{Chen:2015loa,He:2015cea}).
Work has followed in Ref.~\cite{Chen:2015sxa} with the study of the  $\Xi^-_b \to J/\psi K^- \Lambda$ decay, 
suggesting that a strange hidden charm state also predicted in Refs.~\cite{Wu:2010jy,Wu:2010vk} could be seen 
in the $J/\psi  \Lambda$ mass distribution.
%\footnote{This mode is now under present investigation by the LHCb collaboration.}
Interestingly, in Ref.~\cite{Aaij:2014zoa} the LHCb Collaboration had also observed a peak at about the same mass in the  
$J/\psi p$ mass distribution of the $\Lambda_b^0 \rightarrow J/\psi p \pi^-$ reaction, for which no comment was 
done in that paper nor in Refs.~\cite{Aaij:2015tga,Aaij:2015fea} (see \cite{Burns:2015dwa} for 
further comments).
A work along the same lines as \cite{Roca:2015dva} was done for this latter reaction 
in Ref.~\cite{Wang:2015pcn}, showing consistency of the peak seen in the $\Lambda_b^0 \rightarrow J/\psi p \pi^-$ reaction 
with the narrow one observed in the $\Lambda_b \to J/\psi K^-p$ one. 
Very recently, a reanalysis of the experiment of Ref.~\cite{Aaij:2014zoa} has been done by 
the LHCb Collaboration~\cite{Aaij:2016ymb}, concluding that the peak observed in Ref.~\cite{Aaij:2014zoa} is indeed 
consistent with the claims of two states made in Refs.~\cite{Aaij:2015tga,Aaij:2015fea}.
The strange hidden charm state of \cite{Wu:2010jy,Wu:2010vk} was also suggested to be searched for 
in the $J/\psi \Lambda$ mass distribution in the $\Lambda_b^0 \rightarrow J/\psi \eta \Lambda$ reaction in Ref.~\cite{Feijoo:2015kts}, in 
the $\Lambda_b\to J/\psi K^0\Lambda$ reaction in Ref.~\cite{Lu:2016roh} and in 
the $\Xi_b^-\to J/\psi K^-\Lambda$ in Ref.~\cite{Chen:2015sxa}. Discussions on these and other 
reactions can be seen in Refs~\cite{Cheng:2015cca,Chen:2016heh,Chen:2016qju, Oset:2016nvf}.

The search for pentaquark states is not the only relevant information obtained from these reactions. 
Indeed, one of the interesting findings in Ref.~\cite{Roca:2015tea} was that the reaction, 
$\Lambda_b\to J/\psi + \pi\Sigma(\bar{K}N)$, 
acted as a filter for $I=0$ baryon states, which was later confirmed by the analysis of \cite{Aaij:2015tga,Aaij:2015fea}, where only the $\Lambda\ (I=0)$ states were seen in the $K^- p$ mass distribution. This was used \cite{Roca:2015tea} to make predictions for the shape of the $\Lambda(1405)$ in the $\pi \Sigma$ mass distribution. 
Similarly, in Ref.~\cite{Wang:2015pcn} it was shown that the $\Lambda_b\to J/\psi p\pi$ decay was a good filter for baryons with $I=1/2$ and, indeed, one can see in the experiment that there is no trace for the $\Delta(1232)$ which otherwise is present with large strength in most pionic reactions. This filtering of quantum numbers, in spite of the weak interaction not conserving isospin, is tied to rules selecting Cabibbo favored reactions and to dynamical mechanisms that leave the light quarks of the $\Lambda_b$ as spectators in the reaction. These filters make these decays particularly suitable to study baryon resonances that in most reactions appear together with contributions of other isospin channels. Taking advantage of this interesting property the $\Lambda_b \to J/\psi K \Xi$ was suggested ~\cite{Feijoo:2015cca} as a tool to investigate the  $K \Xi$ interaction in the $I=0$ sector.  In a similar way, in Ref.~\cite{Miyahara:2015cja} the $\Lambda_{c}^+ \to \pi+ \pi\Sigma (\bar K N, \eta \Lambda)$ was studied and shown to be also a good filter for $I=0$ baryon states, allowing one to see the $\Lambda(1405)$ and  $\Lambda(1670)$ resonances. 

 In the present work, we take advantage of these ideas and study the $\Xi^+_c \to \pi^+ (\bar K \Sigma, \bar K \Lambda, \pi \Xi)$ reactions, 
 showing that they provide a good filter for $I=1/2$, $S=-2$ resonances,  which thus can be used to learn more about the $\Xi(1620)$ and $\Xi(1690)$ resonances. The $\Xi(1620)$ is cataloged 
 in the Particle Data Group (PDG) with only one star and its spin and parity are unknown~\cite{Agashe:2014kda}. The $\Xi(1690)$ 
 appears there with three stars but its spin-parity quantum numbers are also undetermined. This latter resonance is, on the other hand, located  
 quite close to the $\bar K \Sigma$  threshold, and thus the influence of this threshold on the nature of the $\Xi(1690)$ deserves further 
 study. We shall also see that the $\Xi^+_c \to \pi^+ (\bar K \Sigma, \bar K \Lambda, \pi \Xi)$ decay filters the 
 spin and parity of the final $MB$ pair,  and hence the 
 observation of the $\Xi(1620)$ and $\Xi(1690)$ states in this reaction might allow to determine the unknown spin and 
 parity of these resonances.

%=====================================================================
\section{$\Xi$ resonances}

%----------- exp. ---------------
Although the number of $\Xi$ states should be comparable with that  of nucleon resonances from the viewpoint of quark 
models, at present, the number of measured $\Xi$ states is significantly smaller~\cite{Agashe:2014kda}. Therefore, the study 
of $\Xi$ resonances is relevant in connection with the underlying baryon structure.
The assignment of the spin-parity, $J^P$, in most of the known $\Xi$ resonances is also incomplete, and thus these quantum numbers have been 
determined only for  
few of them: the ground octet $\Xi(1320)$ and 
decuplet $\Xi(1530)$ states and the excited $\Xi(1820)$ resonance. 
The $\Xi(1690)$ is a PDG three-star state, with $(M,\Gamma)=(1690\pm10\ {\rm MeV}, <30\ {\rm MeV})$, 
where $M$ and $\Gamma$ represent the mass and the width, respectively. 
It was first observed in the reaction, $K^-p\to (\bar{K}\Sigma)K\pi$, as a threshold 
enhancement  in the neutral and negatively charged $\bar{K}\Sigma$ mass spectra~\cite{Dionisi:1978tg}. Subsequently, the 
resonance has been also observed in  hyperon-nucleon interactions~\cite{Biagi:1981cu,Biagi:1986zj,Adamovich:1997ud}.
As explained in Sec.~\ref{sec:intro}, recently, heavy hadron decays have begun to emerge as a 
new analysis method for hadron spectroscopy. 
The $\Xi(1690)$ has been studied in some charmed hadron decays like those of the $\Lambda_c$ and $\psi(3686)$ 
hadrons~\cite{Abe:2001mb,Link:2005ut,Aubert:2008ty,Ablikim:2015swa}. In one of such recent experiments,  
$\Lambda_c^+ \to \Xi^-\pi^+K^+$, 
the BaBar Collaboration~\cite{Aubert:2008ty} has 
found some evidence supporting spin-parity quantum numbers $J^P=1/2^-$ for this resonance. The 
spin $J=1/2$ is also favored by the analysis of the $\Lambda_c^+\to\Lambda\bar{K}^0K^+$ reaction~\cite{Aubert:2006ux}.
Nevertheless to fully clarify the $\Xi(1690)$ quantum numbers, further experiments are certainly required. 

%----------  theory -------------
% quark model
In the theoretical side, the description of the $\Xi(1690)$ has been somehow controversial~\cite{Chao:1980em,Capstick:1986bm,Glozman:1995fu,Pervin:2007wa,Melde:2008yr,Xiao:2013xi,Ramos:2002xh,GarciaRecio:2003ks,Gamermann:2011mq,Sekihara:2015qqa,Schat:2001xr,Oh:2007cr}. 
In quark models, 
the difficulty arises in assigning its spin-parity. For example, the nonrelativistic quark model in Ref.~\cite{Chao:1980em} predicted 
the first radial excitation with $J^P=1/2^+$ around 1690 MeV. On the other hand, Ref.~\cite{Pervin:2007wa} assigned the $\Xi(1690)$ to the first 
orbital excitation with $J^P=1/2^-$, and Ref.~\cite{Xiao:2013xi} supported this assignment analyzing  its decay width. In addition, it is also difficult to 
reproduce its mass, and several works predict masses for the $\Xi(1690)$ significantly above  the experimental value~\cite{Capstick:1986bm,Glozman:1995fu,Melde:2008yr}.
% other approach
There are other approaches based on large $N_c$ QCD~\cite{Schat:2001xr} and the Skyrme model~\cite{Oh:2007cr}. The former obtained a $\Xi$ resonance with $J^P=1/2^-$ which has a much larger mass than the $\Xi(1690)$, and the latter predicted two $\Xi$ resonances with $J^P=1/2^-$ which have masses consistent with the experimental values of the $\Xi(1620)$ and $\Xi(1690)$.

% chiral unitary
In late years, the meson-baryon scattering in the strangeness $S=-2$ sector has been also studied in different unitary coupled-channel approaches constrained by QCD chiral symmetry~\cite{Ramos:2002xh,GarciaRecio:2003ks,Gamermann:2011mq,Sekihara:2015qqa}.  The $\Xi(1690)$ was dynamically generated in Refs.~\cite{GarciaRecio:2003ks,Gamermann:2011mq,Sekihara:2015qqa}, and it 
turned to have a quite small  width of only around few MeV. In these schemes, the $\Xi(1690)$ would have spin-parity $J^P=1/2^-$ and it would strongly couple to $\bar{K}\Sigma$ and $\eta\Xi$, having thus large molecular components~\cite{Sekihara:2015qqa}. However, this state did not appear in the analysis of  Ref.~\cite{Ramos:2002xh}, where the authors suggested that the $\Xi(1690)$ might not be a molecular state. In all the chiral unitary approaches~\cite{Ramos:2002xh,GarciaRecio:2003ks,Gamermann:2011mq,Sekihara:2015qqa} the $\Xi(1620)$ is also generated, with a relatively large decay width. This state strongly couples to  $\pi\Xi$ and $\bar{K}\Lambda$, and it is thought to be originated from the strong attraction in the $\pi\Xi$ channel~\cite{GarciaRecio:2003ks,Gamermann:2011mq,Sekihara:2015qqa}. The experimental evidence for the $\Xi(1620)$ is quite poor, and the PDG assigns to this state only one-star~\cite{Agashe:2014kda}. 
Considering such situation, the analysis of these $\Xi$ resonances is interesting, and important for the search of exotic states, which are not easily accommodated as three-body quark states. 
%\textcolor{red}{(Search for results of Lattice and Skyrme model)}

%------------ this work ---------------
In this work, to study $\Xi(1690)$ and $\Xi(1620)$ we  analyze the $\Xi_c\to\pi^+(MB)_i$ decay ($M$ and $B$ represent the meson and baryon, respectively, with the index $i$ denoting the meson-baryon channel). To account for the final meson-baryon interaction, we examine the predictions deduced from the chiral unitary 
approaches of Refs.~\cite{Ramos:2002xh,GarciaRecio:2003ks,Sekihara:2015qqa}. 

\begin{figure}[tb]
\begin{center}
\includegraphics[width=8cm,bb=0 0 988 722]{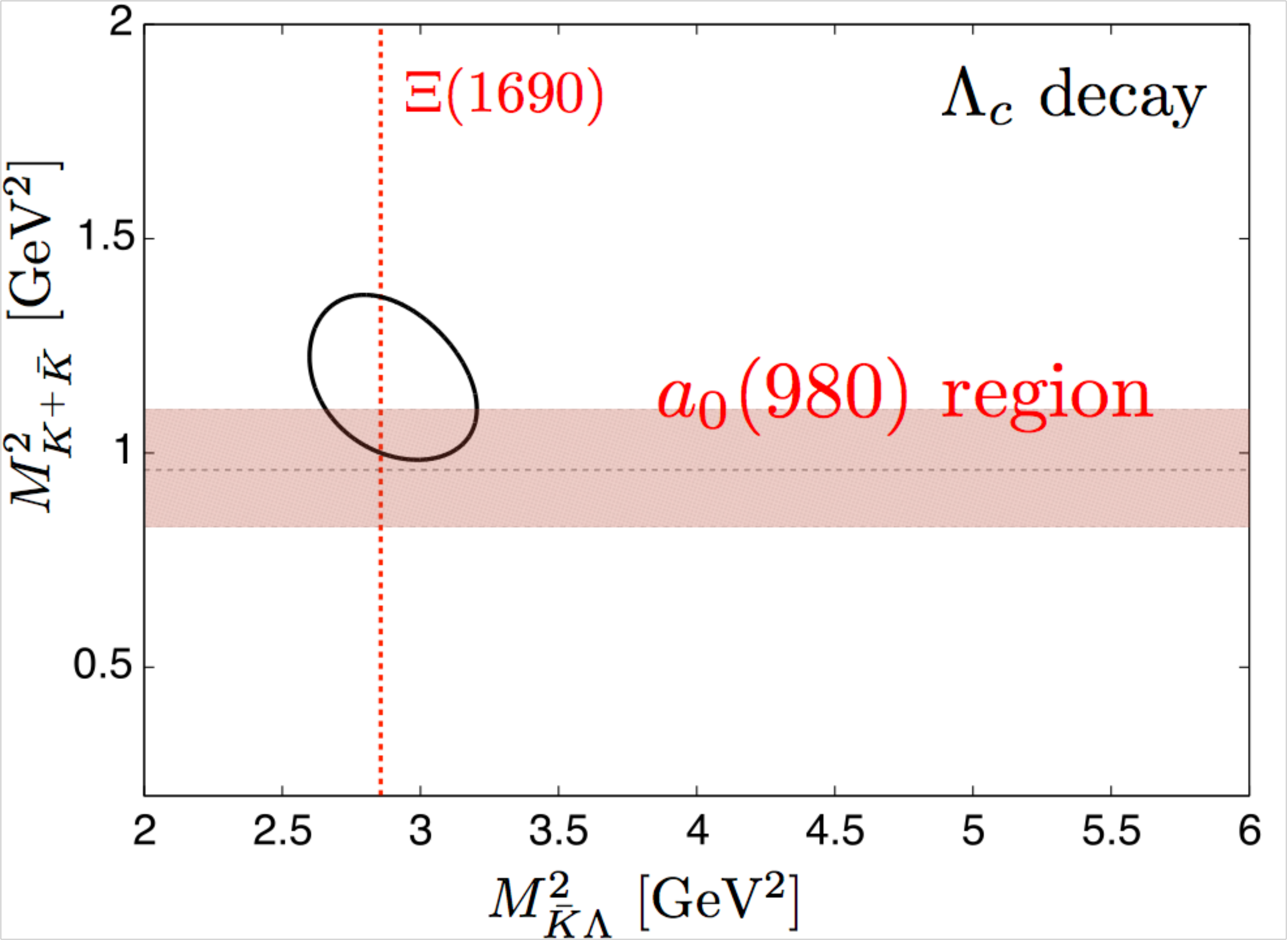}
\caption{$M^2_{\bar{K}\Lambda}$ and $M^2_{K^+\bar{K}}$ Dalitz plot for the  $\Lambda_c\to K^+(\bar{K}\Lambda)$ reaction. The $\Xi(1690)$ energy is shown by the vertical dotted line, while the horizontal band represents the mass and the width of the $a_0(980)$.}
\label{fig:dalitz1}
%\ref{fig:dalitz1}
\end{center}
\end{figure}

\begin{figure}[tb]
\begin{center}
\includegraphics[width=8cm,bb=0 0 988 722]{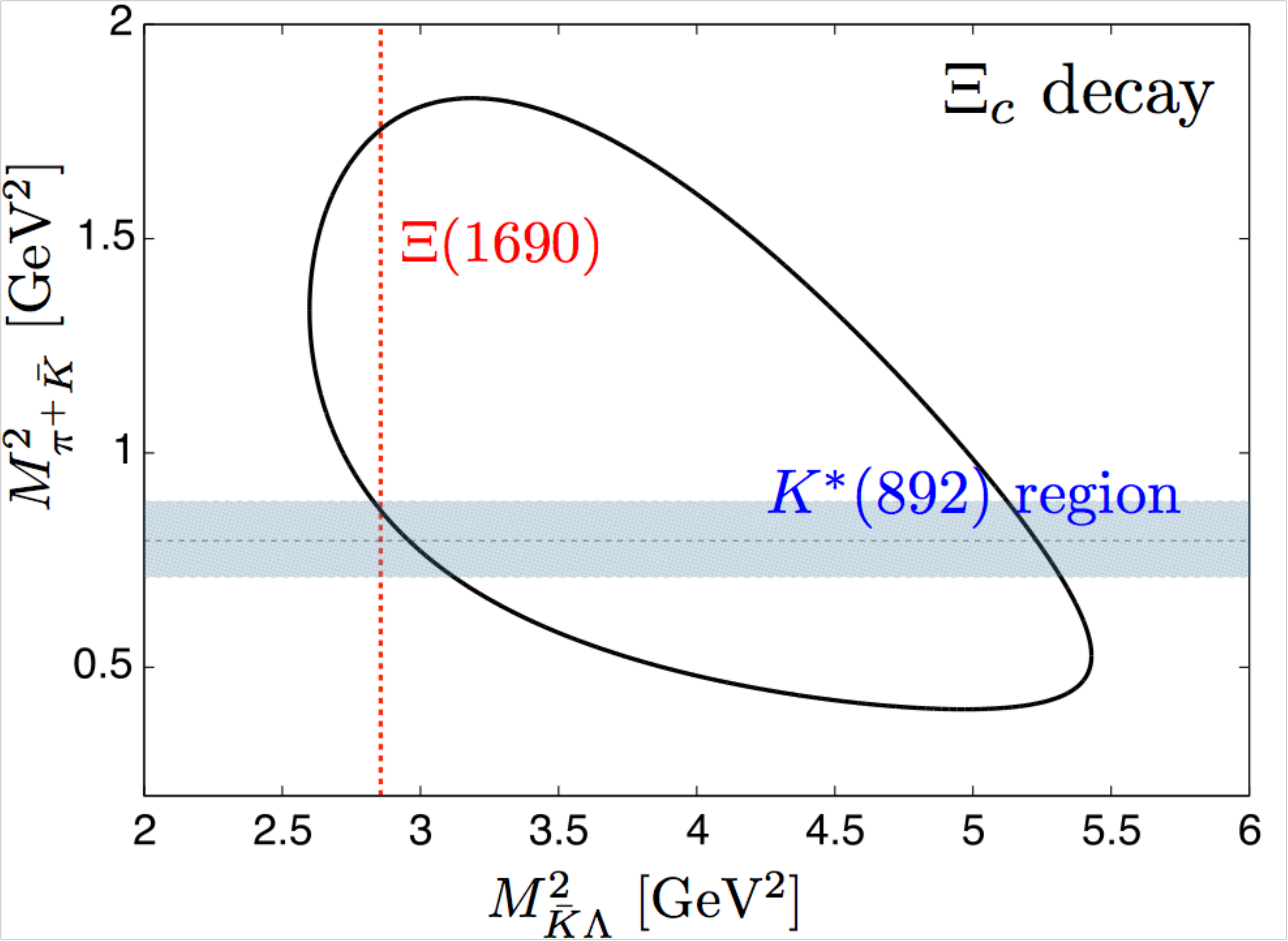}
\caption{$M^2_{\bar{K}\Lambda}$ and $M^2_{\pi^+\bar{K}}$ Dalitz plot  for the  $\Xi_c\to \pi^+(\bar{K}\Lambda)$ reaction. As in Fig.~\ref{fig:dalitz1}, the $\Xi(1690)$ energy is shown by the vertical dotted line. The horizontal band represents the mass and the width of the $K^*(892)$.}
\label{fig:dalitz2}
%\ref{fig:dalitz2}
\end{center}
\end{figure}

As mentioned above, experimentally the $\Lambda_c$ decay has been also examined to extract information about the $\Xi$ states. 
We compare both, $\Lambda_c$ and $\Xi_c$, decay reactions,  from the viewpoint of the kinematics and we show
Dalitz plots of the $\Xi_c\to \pi^+(\bar{K}\Lambda)$ and $\Lambda_c\to K^+(\bar{K}\Lambda)$ reactions in Figs.~\ref{fig:dalitz1} and ~\ref{fig:dalitz2}, respectively.
%%%%%%% figure  %%%%%%%%%
%%%%%%%%%%%%%%%%%%%%
%%%%%%%%%%%%%%%%%%%%
%%%%%%%%%%%%%%%%%%%%
%
%%%%%%% figure  %%%%%%%%%
%%%%%%%%%%%%%%%%%%%%
%%%%%%%%%%%%%%%%%%%%1
%%%%%%%%%%%%%%%%%%%%
In the $\Xi(1690)$ energy region, the $\Lambda_c$ decay Dalitz plot overlaps greatly with the $a_0(980)$ meson resonance in the $K\bar{K}$ channel, which makes the $\Xi(1690)$ analysis difficult~\cite{Aubert:2006ux}. On the other hand, in the $\Xi_c$ decay, the overlap with the corresponding meson resonance, the $K^*(892)$ now in the $\pi\bar{K}$ channel, is much smaller. Furthermore, if we choose the $\Xi_c^+\to\pi^+(\bar{K}^0\Lambda)$ reaction  instead of the $\Xi_c^0\to\pi^+(K^-\Lambda)$ decay, the $\pi^+\bar{K}^0$ pair must be in an isospin  $I=3/2$ state, since its third component is $+3/2$. This means that the analysis of the $\Xi_c^+$ decay 
should not be influenced by the presence of meson resonances, since  an isospin $I=3/2$ meson would be certainly an exotic state. Hence, the analysis of the $\Xi_c$ decays, in particular that of the  $\Xi_c^+$, is an ideal reaction for the study of strangeness $\Xi$ baryons. 
There exist several excited baryon resonances in the $\bar{K}\Lambda$ and $\pi\Lambda$ channels around the $\Xi(1690)$ energy region. However, because such resonances are quite broad and their large overlap, 
it is reasonable to suppose that their corresponding bands would not be visible in the Dalitz plot in sharp contrast to the  meson resonance cases discussed above.

%=====================================================================
\section{Formulation}    \label{sec:formulation}  %\ref{sec:formulation}
\begin{figure}[tb]
\begin{center}
\includegraphics[width=8cm,bb=70 0 890 366]{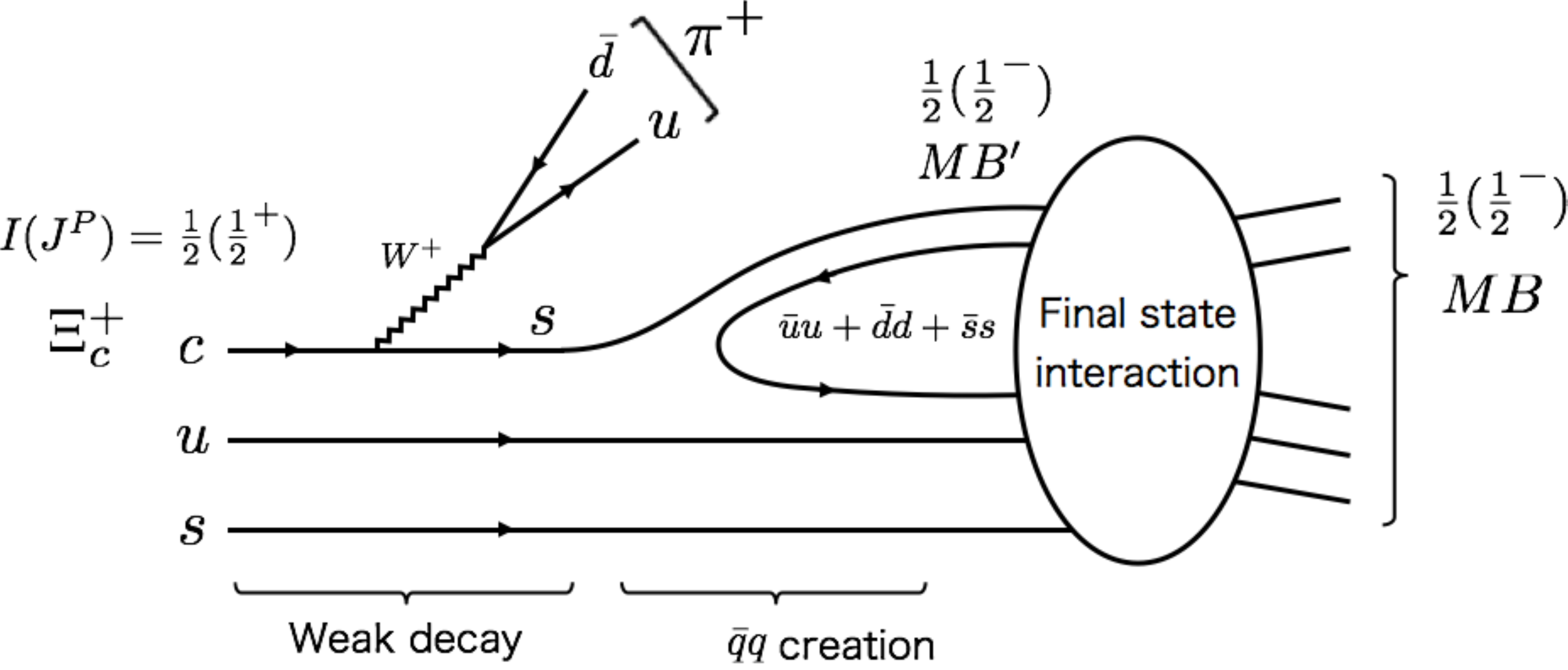}
\caption{Dominant quark-line diagram for the $\Xi_c^+\to \pi^+MB$ decay. The solid and the wiggly lines stand for the quarks and the $W$ boson, respectively. }
\label{fig:Xic_decay}  %\ref{fig:Xic_decay}
\end{center}
\end{figure}

Following our previous work~\cite{Miyahara:2015cja}, we show in Fig.~\ref{fig:Xic_decay} the dominant quark-line diagram for the  $\Xi_c^+\to \pi^+(MB)$ decay, when the final $MB$ pair is emitted close to threshold. 
%%%%%%%%%%%  figure  %%%%%%%%%%%%
%%%%%%%%%%%%%%%%%%%%%%%%%%%
%%%%%%%%%%%%%%%%%%%%%%%%%%%
%%%%%%%%%%%%%%%%%%%%%%%%%%%
%
We split the decay process in three parts. The first one involves the $c \to s$ weak transition and the production of a high momentum $\pi^+$. Next we consider the $\bar{q}q$ creation part, where the intermediate meson-baryon states are constructed with certain weights. Finally, we have the rescattering of the intermediate meson-baryon pairs which will be taken into account in a coupled channel chiral unitary scheme. 

In what follows, we will focus on the $\Xi_c^+$ decay. The analysis of the  $\Xi_c^0$ decay runs in parallel, because the dominant quark-line diagram is similar to that shown in Fig.~\ref{fig:Xic_decay}. There exist however some differences induced by subdominant mechanisms, which will be discussed in Sec.~\ref{sec:Xic0}.

%-------------------------------------------------------------------------------------------------------------------------
\subsection{Weak decay}  \label{sec:weak}  %\ref{sec:weak}
\begin{figure}[tb]
\begin{center}
\includegraphics[width=8cm,bb=-180 0 472 172]{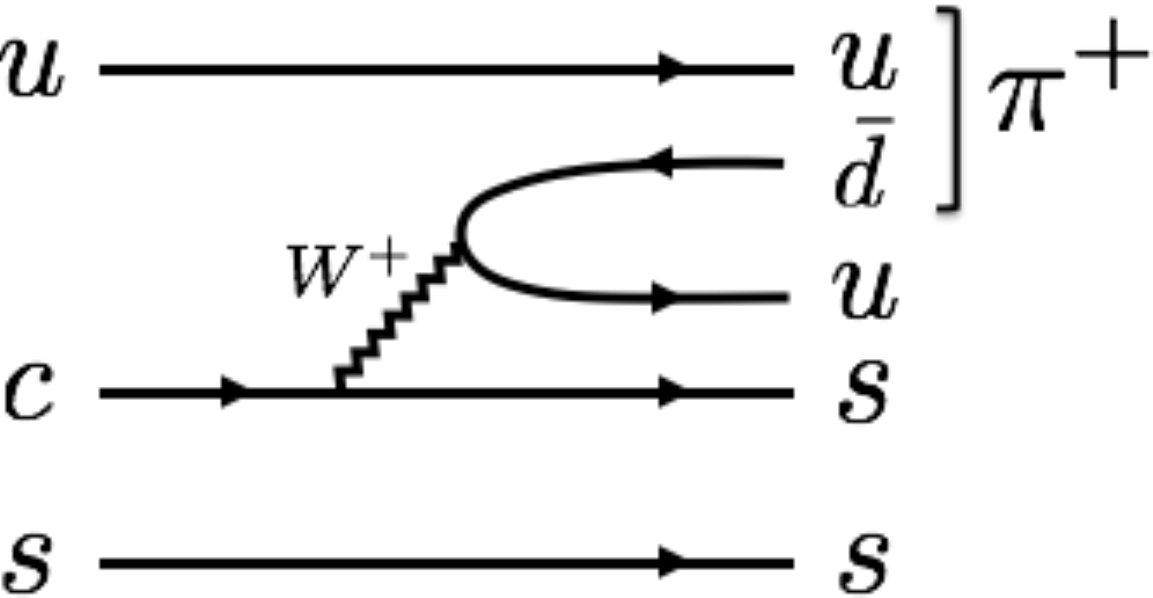}
\caption{Subdominant mechanism for the $\Xi_c^+\to \pi^+MB$ decay. Though its contribution is also Cabibbo allowed, it is however suppressed when compared to that depicted in  Fig.~\ref{fig:Xic_decay} (see text for details).  }
\label{fig:Xic_decay2}  %\ref{fig:Xic_decay2}
\end{center}
\end{figure}

The Cabibbo allowed reactions of interest for the $\Xi_c$ decay  are $c\to su\bar{d}$ and $cd\to su$. When it is  required the emission of high momentum $\pi^+$,  these reactions  lead to the two quark-line diagrams depicted in Figs.~\ref{fig:Xic_decay} and \ref{fig:Xic_decay2}, respectively.
%%%%%%%%%%%  figure  %%%%%%%%%%%%
%%%%%%%%%%%%%%%%%%%%%%%%%%%
%%%%%%%%%%%%%%%%%%%%%%%%%%%
%%%%%%%%%%%%%%%%%%%%%%%%%%%
%
However, the mechanism in Fig.~\ref{fig:Xic_decay2} is suppressed in comparison with that shown in Fig.~\ref{fig:Xic_decay}.  First there is a color enhancement factor in the latter one, which is not present in the diagram of   Fig.~\ref{fig:Xic_decay2}. This is because in the $W$ boson-$u\bar{d}$ vertex, the  color of the outgoing quarks is fixed by that of the $u$ quark belonging to  the $\Xi_c$ since a  color singlet ($\pi^+$) needs to be constructed. In contrast, in  the mechanism of Fig.~\ref{fig:Xic_decay} all the colors are allowed in the $W$ vertex. On the other hand, the $u$ and $s$ quarks in the $\Xi_c$ form a strongly correlated antisymmetric diquark configuration difficult to separate. Therefore, a mechanism where the diquark state is destroyed like that  depicted in Fig.~\ref{fig:Xic_decay2} is expected to be suppressed. Finally kinematics also favor the diagram of Fig.~\ref{fig:Xic_decay} since we will be interested in situations where the outgoing $MB$ pair is produced at low invariant masses (see Fig.~\ref{fig:dalitz2}), which in turn requires  the emission of a high momentum $\pi^+$. Because the $u$ quark in the $\Xi_c$ is a spectator in Fig.~\ref{fig:Xic_decay2},  it is at rest in the $\Xi_c$ center of mass frame and thus, it is difficult its association with a high momentum $\bar{d}$, coming from the $W$ decay, to construct the final high energy $\pi^+$. 

For all the above arguments, we think the mechanism depicted in Fig.~\ref{fig:Xic_decay} should be dominant in the  $\Xi_c$ decay, and we will use it to study the influence of the $\Xi$ resonances in the process. Attending to the structure of the quark degrees of freedom, the ground state of the $\Xi_c$ is almost dominated by the flavor SU(3)-subgroup ${\bf \bar{3}}$ configuration~\cite{Roberts:2007ni}, 
\begin{align}
|\Xi_c\rangle = \frac{1}{\sqrt{2}}|c(su-us)\rangle. %\notag
\end{align}
Therefore, the $sus$ cluster formed after the charm quark decay will be
\begin{align}
\frac{1}{\sqrt{2}}|s(su-us)\rangle.  %\notag
\end{align}

%-------------------------------------------------------------------------------------------------------------------------
\subsection{$\bar{q}q$ creation}
The next step is the insertion of the vacuum-quantum-numbers $\bar{q}q$-pair creation to construct the intermediate meson-baryon state ${MB}^\prime$. To analyze decay modes where final $MB$ state has  $J^P=1/2^-$ spin-parity, the $s$ quark originated in the weak decay should carry one unit of angular momentum, $L=1$. On the other hand, assuming ground states  and a relative $s$ wave for the final $MB$, the $\bar{q}q$ creation should be attached precisely to this $s$ quark. We further assume that the $u$ and $s$ quarks belonging to the $\Xi_c$ baryon and spectators in the decay in the mechanism of Fig.~\ref{fig:Xic_decay}, keep the strong diquark correlation discussed in the previous section. Hence, after the $\bar{q}q$ creation, these $u$ and $d$ quarks should be part of the baryon, and the $s$ quark originated in the weak decay should form the meson, as shown in Fig.~\ref{fig:Xic_decay}. The above picture leads to
\begin{align}
|MB^\prime\rangle &= \frac{1}{\sqrt{2}}|s(\bar{u}u+\bar{d}d+\bar{s}s)(su-us)\rangle \label{eq:fraction_quark}
%\eqref{eq:fraction_quark}
%&=\frac{1}{\sqrt{2}}\sum_{i=1}^{3} |P_{3i}q_i(su-us)\rangle, %\notag
\end{align}
%
%where
%\begin{align}
%q&\equiv
%\left( \begin{array}{c}
%u \\ d \\ s
%\end{array} \right), \notag \\
%P&\equiv q\bar{q}=
%\left( \begin{array}{ccc}
%u\bar{u}  &  u\bar{d}  &  u\bar{s}  \\
%d\bar{u}  &  d\bar{d}  &  d\bar{s}  \\
%s\bar{u}  &  s\bar{d}  &  s\bar{s}  
%\end{array} \right).  %\notag
%\end{align}
%%
%$P$ can be written in terms of mesonic degrees of freedom,
%%
%\begin{align}
%P=\left( \begin{array}{ccc}
%\frac{\pi^0}{\sqrt{2}}+\frac{\eta}{\sqrt{3}}+\frac{\eta^\prime}{\sqrt{6}}  &  \pi^+  &  K^+  \\
%\pi^-  &  -\frac{\pi^0}{\sqrt{2}}+\frac{\eta}{\sqrt{3}}+\frac{\eta^\prime}{\sqrt{6}}  &  K^0  \\
%K^-  &  \bar{K}^0  &  -\frac{\eta}{\sqrt{3}}+\frac{2\eta^\prime}{\sqrt{6}}
%\end{array} \right).  %\notag
%\end{align}
%The mixed antisymmetric flavor state of baryons reads~\cite{Close:1979bt}
%\begin{align}
%|\Sigma^+\rangle &= -\frac{1}{\sqrt{2}}|uus-usu\rangle,  \notag \\
%|\Sigma^0\rangle &= \frac{1}{2}|uds+dus-dsu-usd\rangle, \notag \\
%|\Lambda\rangle &= \frac{1}{\sqrt{12}}|usd-dsu+dus-uds-2sdu+2sud\rangle, \notag \\
%|\Xi^0\rangle &= \frac{1}{\sqrt{2}}|sus-ssu\rangle. %\notag
%\end{align}
%%
%Using these hadron degrees of freedom, we can rewrite the intermediate state as
As explained in the Appendix, we can connect two degrees of freedom, the quarks and the hadrons. 
Using the quark representations of hadrons discussed in Appendix, we can rewrite the intermediate state as
\begin{align}
|MB^\prime\rangle &=|K^-\Sigma^+\rangle -\frac{1}{\sqrt{2}}|\bar{K}^0\Sigma^0\rangle +\frac{1}{\sqrt{6}}|\bar{K}^0\Lambda\rangle -\frac{1}{\sqrt{3}}|\eta\Xi^0\rangle.  \label{eq:MB}
%\eqref{eq:MB}
\end{align}
In the isospin basis, this becomes%%% footnote
\footnote{We follow the convention of Ref.~\cite{Oset:1998it}:
%%%
\begin{align}
|\pi^+\rangle&=-|I=1,I_z=1\rangle,  \notag \\
|K^-\rangle&=-|1/2,-1/2\rangle, \notag \\
|\Sigma^+\rangle&=-|1,1\rangle, \notag \\
|\Xi^-\rangle&=-|1/2,-1/2\rangle. \notag
\end{align}} 
%%% end footnote
%
\begin{align}
|MB^\prime\rangle = \frac{1}{\sqrt{6}}|\bar{K}\Lambda\rangle -\sqrt{\frac{3}{2}}|\bar{K}\Sigma\rangle -\frac{1}{\sqrt{3}}|\eta\Xi\rangle,  \label{eq:MB_iso}
%\eqref{eq:MB_iso}
\end{align}
where the isospin quantum numbers of all states are  $I=1/2$. In Eqs.~\eqref{eq:MB} and \eqref{eq:MB_iso}, we have neglected the contribution from the $\eta^\prime\Xi$ channel because its threshold is located  much higher in energy. 

\begin{figure}[tb]
\begin{center}
\includegraphics[width=8cm,bb=0 0 744 626]{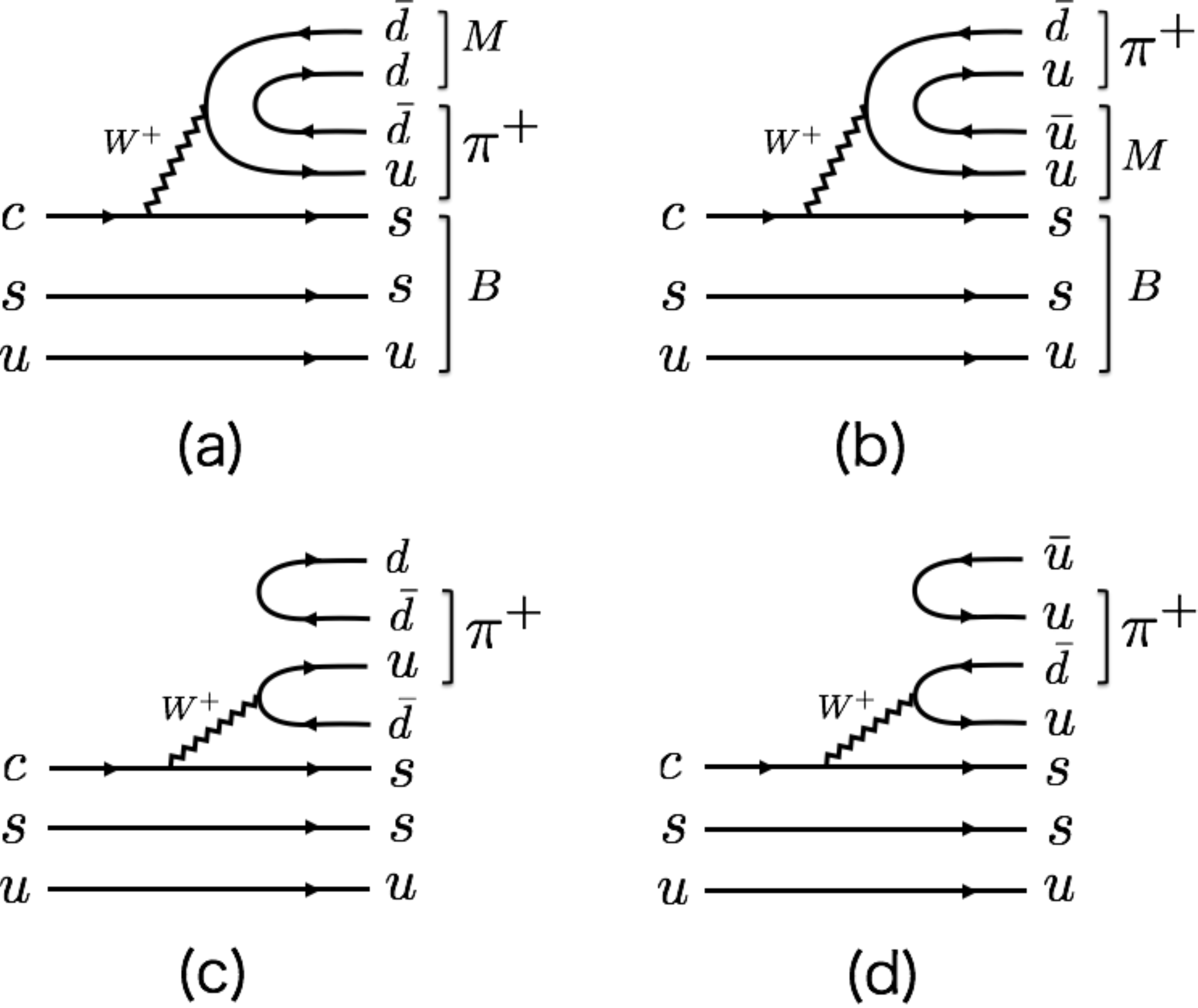}
\caption{Other possible mechanisms for the $\Xi_c^+\to \pi^+MB$ decay. }
\label{fig:Xic_decay3}  %\ref{fig:Xic_decay3}
\end{center}
\end{figure}

%------ other diagram -------
Up to this point, we have considered mechanisms where the high momentum $\pi^+$ is emitted right after the weak $c\to s$ transition, and its formation is independent of the vacuum quark-antiquark pair creation.  
It is natural to consider these quark-line diagrams first, because we assume the emitted pion has relatively a large momentum so that the remaining system is close to the meson-baryon thresholds. Such diagram approach is known to be (qualitatively) powerful in the hadronic weak decays.
However, there are other quark-line diagrams where the $\pi^+$ is emitted after the $\bar{q}q$ insertion, as shown in Fig.~\ref{fig:Xic_decay3}.
%%%%%%%%%%%  figure  %%%%%%%%%%%%
%%%%%%%%%%%%%%%%%%%%%%%%%%%
%%%%%%%%%%%%%%%%%%%%%%%%%%%
%%%%%%%%%%%%%%%%%%%%%%%%%%%
Although a momentum mismatch will suppress the emission of high-momentum pion for the soft $\bar{q}q$ pair creation, the contributions of such processes might be non-negligible.
The quark-line diagrams Figs.~\ref{fig:Xic_decay3}(c) and \ref{fig:Xic_decay3}(d) are suppressed because of the color recombination factors, similarly as it was discussed above. 
Indeed, the color of the $\bar{d}$ or $u$ quarks, respectively, from the $c$ weak decay is fixed since 
it should be coupled to the quarks in the $ssu$ cluster. 
However, there are no robust reasons to exclude the contribution from the mechanisms depicted in Figs.~\ref{fig:Xic_decay3}(a) and \ref{fig:Xic_decay3}(b), 
except for the kinematical suppression produced by having a high energy quark emitted from a weak vertex
part of the low energy $MB$ final pair. 
Using the same procedure as above, we obtain  that the intermediate $MB''$ state for these additional diagrams would be
\begin{align}
|MB''\rangle = \frac{2}{\sqrt{3}}|\eta\Xi^0\rangle.  \label{eq:MB_add}
%\eqref{eq:MB_add}
\end{align}
Because we do not specify the detailed mechanism for the $\bar{q}q$ pair creation, the relative phase between the $|MB'\rangle$ and $|MB''\rangle$ intermediate states cannot be determined. We thus introduce a linear combination,
\begin{align}
|MB'\rangle+x|MB''\rangle   \label{eq:MB_add}
%\eqref{eq:MB_add}
\end{align}
with an unknown weight factor $x$. As we will show in Sec.~\ref{sec:other_diagram}, the qualitative features of the spectra are not significantly  affected when values of $x$ in the  $[-1,1]$ range are considered. For the sake of brevity, in what follows we will mainly show results for $x=0$, unless it is otherwise stated.

%-------------------------------------------------------------------------------------------------------------------------
\subsection{Final-state interaction}
\begin{figure}[tb]
\begin{center}
\includegraphics[width=8cm,bb=0 0 911 169]{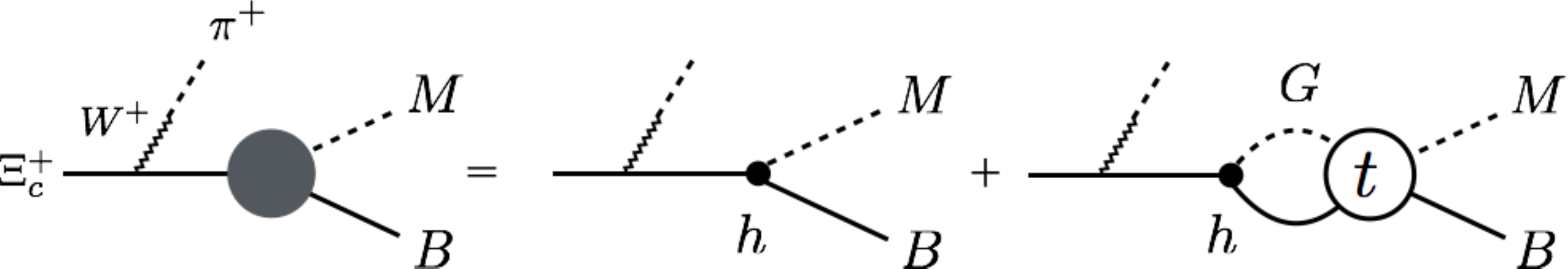}
\caption{Schematic diagram for the FSI of the meson-baryon pair (solid circle). The first and the second terms in the right-hand side stand for the  tree and the rescattering contributions, respectively. The latter diagram contains the meson-baryon loop function $G$ and the scattering amplitude $t$. Besides, the factor $h$ represents the intermediate meson-baryon weight  introduced in Eqs.~\eqref{eq:MB} or \eqref{eq:MB_iso}.}
\label{fig:FSI}  %\ref{fig:FSI}
\end{center}
\end{figure}

The intermediate mesons and baryons [Eqs.~\eqref{eq:MB} or \eqref{eq:MB_iso}] re-scatter through strong interactions, and produce the decay amplitude $\mathscr{M}_j$ for the final meson-baryon $MB$ pair. The schematic diagram is shown in Fig.~\ref{fig:FSI}, 
%%%%%%%%%%%  figure  %%%%%%%%%%%%
%%%%%%%%%%%%%%%%%%%%%%%%%%%
%%%%%%%%%%%%%%%%%%%%%%%%%%%
%%%%%%%%%%%%%%%%%%%%%%%%%%%
where the total contribution is the sum of the tree diagram obtained directly from the $\bar{q}q$ creation, and the rescattering term that accounts for the final-state interaction (FSI) of the intermediate meson-baryon pairs.  The factor $h$ represents the coefficients in Eqs.~\eqref{eq:MB} or ~\eqref{eq:MB_iso}.
The meson-baryon loop function $G$ and the meson-baryon scattering amplitude $t$ are calculated using a certain model. In this work, we use some chiral unitary approaches as explained in Sec.~\ref{results}.

The decay amplitude $\mathscr{M}_j$ to the final meson-baryon state $MB$ is expressed as
\begin{align}
\mathscr{M}_j =V_P &\left( h_j + \sum_i h_i G_i(M_{\rm inv})t_{ij}(M_{\rm inv}) \right),  \label{eq:amplitude} 
%\eqref{eq:amplitude}
\end{align}
where $M_{\rm inv}$ represents the meson-baryon invariant mass. The dynamics before the FSI is all included in $V_P$, which is assumed to be constant in the relevant energy region. The actual value of $V_P$ may be determined from an experimental measurement of the decay distribution in a certain decay channel. 
The coefficients $h_j$ for the physical basis are
\begin{align}
h_{\pi^0\Xi^0}&=h_{\pi^+\Xi^-}=0,\ h_{\bar{K}^0\Lambda}=\frac{1}{\sqrt{6}},  \notag \\
h_{K^-\Sigma^+}&=1,\ h_{\bar{K}^0\Sigma^0}=-\frac{1}{\sqrt{2}},\ h_{\eta\Xi^0}=-\frac{1}{\sqrt{3}},  %\notag
\end{align}
and for the isospin basis are
\begin{align}
h_{\pi\Xi}&=0,\ 
h_{\bar{K}\Lambda}=\frac{1}{\sqrt{6}},\ 
h_{\bar{K}\Sigma}=-\sqrt{\frac{3}{2}},\ 
h_{\eta\Xi}=-\frac{1}{\sqrt{3}}.  %\notag
\end{align}
The reason for the vanishing $\pi\Xi$ coefficients can be understood from the decay mechanism as shown in Fig.~\ref{fig:Xic_decay}, where the meson $M$ consists of the $s$ quark and cannot be the $\pi$ after the $\bar{q}q$ creation.
For nonzero $x$ in Eq.~\eqref{eq:MB_add}, $h_{\eta\Xi}$ is modified as 
$h_{\eta\Xi}=(-1+2x)/\sqrt{3}$.
To directly compare with the experimental data, we rewrite the amplitude in the isospin basis;
\begin{align}
\mathscr{M}_{\pi^0\Xi^0} &= -\frac{1}{\sqrt{3}}\mathscr{M}_{\pi\Xi}^{I=1/2}+\sqrt{\frac{2}{3}}\mathscr{M}_{\pi\Xi}^{I=3/2},   \notag \\
\mathscr{M}_{\pi^+\Xi^-} &= \sqrt{\frac{2}{3}}\mathscr{M}_{\pi\Xi}^{I=1/2}+\frac{1}{\sqrt{3}}\mathscr{M}_{\pi\Xi}^{I=3/2},   \notag \\
\mathscr{M}_{\bar{K}^0\Lambda} &= \mathscr{M}_{\bar{K}\Lambda}^{I=1/2},  \notag \\
\mathscr{M}_{K^-\Sigma^+} &=  -\sqrt{\frac{2}{3}}\mathscr{M}_{\bar{K}\Sigma}^{I=1/2} +\frac{1}{\sqrt{3}}\mathscr{M}_{\bar{K}\Sigma}^{I=3/2},  \notag \\
\mathscr{M}_{\bar{K}^0\Sigma^0} &= \frac{1}{\sqrt{3}}\mathscr{M}_{\bar{K}\Sigma}^{I=1/2}+\sqrt{\frac{2}{3}}\mathscr{M}_{\bar{K}\Sigma}^{I=3/2},   \notag \\
\mathscr{M}_{\eta\Xi^0} &= \mathscr{M}_{\eta\Xi}^{I=1/2}.  %\notag
\end{align}
Because the $u$ quark in Fig.~\ref{fig:Xic_decay} is a spectator in the weak decay, the final meson-baryon state retains the same isospin as the $\Xi_c$, and the $I=3/2$ sector does not contribute in the decay. Hence, the amplitude can be simplified as 
\begin{align}
\mathscr{M}_{\pi^0\Xi^0} = -\frac{1}{\sqrt{3}}\mathscr{M}_{\pi\Xi}^{I=1/2}&,\ \mathscr{M}_{\pi^+\Xi^-} = \sqrt{\frac{2}{3}}\mathscr{M}_{\pi\Xi}^{I=1/2},   \notag \\
\mathscr{M}_{\bar{K}^0\Lambda} &= \mathscr{M}_{\bar{K}\Lambda}^{I=1/2},  \notag \\
\mathscr{M}_{K^-\Sigma^+} =  -\sqrt{\frac{2}{3}}\mathscr{M}_{\bar{K}\Sigma}^{I=1/2}&,\ \mathscr{M}_{\bar{K}^0\Sigma^0} = \frac{1}{\sqrt{3}}\mathscr{M}_{\bar{K}\Sigma}^{I=1/2},   \notag \\
\mathscr{M}_{\eta\Xi^0} &= \mathscr{M}_{\eta\Xi}^{I=1/2}.  %\notag
\end{align}
With the above decay amplitudes, we can calculate the partial decay width $\Gamma_j$,
\begin{align}
\Gamma_j &=\int d\Pi_3 |\mathscr{M}_j|^2,  \label{eq:width}
%\eqref{eq:width}
\end{align}
where $d\Pi_3$ represents the three-body phase space. The invariant mass distribution is obtained by differentiating the width by $M_{\rm inv}$.

%=====================================================================
\section{Results with chiral unitary approaches}   \label{results}  %\ref{results}

In this section, we use the chiral unitary approaches as the final-state interaction, and show our predictions for different meson-baryon invariant mass distributions in the $\Xi_c$ decay. 

%---------- Ch-U -------------
To quantify systematic uncertainties, we will consider here three chiral unitary approaches, that we will denote by ROB, GLN, and Sekihara, and whose details and predictions can be found in Refs.~\cite{Ramos:2002xh}, \cite{GarciaRecio:2003ks}, and \cite{Sekihara:2015qqa}, respectively.%
\footnote{In Refs.~\cite{Ramos:2002xh,Sekihara:2015qqa}, several parameter sets are introduced. Here, we choose ``Set~5" for the ROB model  and that denoted by ``Fit'' in the Sekihara one.}
 The ROB and GLN approaches are formulated in the isospin symmetric limit, while the Sekihara model uses  physical hadron masses, thus, including some isospin symmetry breaking corrections. 
In Tables~\ref{tab:ROB_GLN_model} and \ref{tab:Sekihara_model}, we compile  the pole positions  and couplings $g_i$ to each $MB$ channel of the resonances found in these references.%
\footnote{In Table~\ref{tab:ROB_GLN_model}, the values of the pole positions and the couplings are slightly different from the ones in the original papers~\cite{Ramos:2002xh,GarciaRecio:2003ks}. 
This is because some small differences in the employed meson and baryon masses.}
%%%%%%%  table   %%%%%%%%%
%%%%%%%%%%%%%%%%%%%%
\begin{table*}[tb]
\begin{center}
\begin{ruledtabular}
{\tabcolsep = 2.5mm 
\begin{tabular}{ccccccc}
  &  &  pole [MeV] &  $g_{\pi\Xi}$ & $g_{\bar{K}\Lambda}$  & $g_{\bar{K}\Sigma}$ & $g_{\eta\Xi}$  \\ 
\hline
ROB (Set~5)~\cite{Ramos:2002xh} & $\Xi(1620)$ & $1606-66i$ & $2.2-0.5i$  & $2.5+0.1i$  & $0.9-0.2i$  & $0.4+0.2i$  \\
\hline
\multirow{2}{*}{GLN~\cite{GarciaRecio:2003ks}}  & $\Xi(1620)$ & $1568-126i$ & $2.2-1.6i$ & $2.2-0.6i$  & $0.7-0.4i$ & $0.1-0.5i$  \\
 & $\Xi(1690)$ & $1667-2i$ & $0.2-0.0i$ & $0.4-0.1i$ & $2.3+0.0i$ & $1.5+0.1i$   \\
\end{tabular}  }
\caption{Pole positions and couplings $g_i$  for the $\Xi$ resonances found in the ROB~\cite{Ramos:2002xh} and the 
GLN~\cite{GarciaRecio:2003ks} models.}
\label{tab:ROB_GLN_model}
%\ref{tab:ROB_GLN_model}
\end{ruledtabular}
\end{center}
\end{table*}
%%%%%%%%%%%%%%%%%%%%
%%%%%%%%%%%%%%%%%%%%
%
%%%%%%%  table   %%%%%%%%%
%%%%%%%%%%%%%%%%%%%%
\begin{table*}[tb]
\begin{center}
\begin{ruledtabular}
%{\tabcolsep = 2.5mm 
\begin{tabular}{ccccccccc}
  &  &  pole [MeV] &  $g_{\pi^0\Xi^0}$ & $g_{\pi^+\Xi^-}$  & $g_{\bar{K}^0\Lambda}$ & $g_{K^-\Sigma^+}$ & $g_{\bar{K}^0\Sigma^0}$ & $g_{\eta\Xi^0}$  \\ 
\hline
Sekihara (Fit)~\cite{Sekihara:2015qqa} & $\Xi(1690)$ & $1684-i$ & $-0.1+0.0i$ & $0.1-0.0i$ & $0.4+0.2i$ & $1.0+0.6i$ & $-0.8-0.4i$ & $-0.7-0.5i$   \\
\end{tabular}  %}
\caption{Pole positions and couplings $g_i$ for the $\Xi$ resonances found in the Sekihara model~\cite{Sekihara:2015qqa}. The $\Xi(1690)$ pole appears in a nonphysical Riemann sheet, above the $K^-\Sigma^+$ threshold.}
\label{tab:Sekihara_model}
%\ref{tab:Sekihara_model}
\end{ruledtabular}
\end{center}
\end{table*}
%%%%%%%%%%%%%%%%%%%%
%%%%%%%%%%%%%%%%%%%%
The poles are found in the appropriate  Riemann sheets defined by continuity with the real axis except for the case of the $\Xi(1690)$ in the Sekihara model, which is found in a nonphysical Riemann sheet above, but quite close to, the $K^-\Sigma^+$ threshold. The $\Xi(1620)$ is dynamically generated in the ROB and GLN models, with large couplings to the $\pi\Xi$ and $\bar{K}\Lambda$ channels. On the other hand, the $\Xi(1690)$ is found in the GLN and Sekihara approaches, with now large couplings to the $\bar{K}\Sigma$ and $\eta\Xi$ channels, but not in the ROB model.

%--------- spectrum ---------------
In the above chiral unitary approaches, only the $s$-wave scattering for $MB$ is considered. In this case, the decay amplitude $\mathscr{M}_j$ depends only on $M_{\rm inv}$ and the invariant mass distributions ${\rm d}\Gamma_j/{\rm d}M_{\rm inv}$ is reduced to
\begin{align}
\frac{{\rm d}\Gamma_j}{{\rm d}M_{\rm inv}} &=\frac{1}{(2\pi)^3}\frac{p_{\pi^+}\tilde{p}_jM_j}{M_{\Xi_c^+}} |\mathscr{M}_j|^2,  \label{eq:massdis}
%\eqref{eq:massdis}
\end{align}
where $M_j$ is the baryon mass in the channel $j$, and $p_{\pi^+}$ ($\tilde{p}_j$) represents the three-momentum of the $\pi^+$  emitted in the weak decay part  (meson in the final $MB$ state)  
in the $\Xi_c$ rest frame (in the $MB$ rest frame),
\begin{align}
p_{\pi^+}=& \frac{\lambda^{1/2}(M_{\Xi_c^+}^2,m_{\pi^+}^2,M_{\rm inv}^2)}{2M_{\Xi_c^+}},\ \tilde{p}_j= \frac{\lambda^{1/2}(M_{\rm inv}^2,M_j^2,m_j^2)}{2M_{\rm inv}},  \notag \\
&\lambda(x,y,z) = x^{2}+y^{2}+z^{2}-2xy-2yz-2zx.  %\notag
\end{align}

As the results of invariant mass distributions, first, we will consider the $\pi\Xi$ channel, which couples strongly to the $\Xi(1620)$ resonance in the ROB and GLN approaches. Later in this section, we will pay attention to 
the $\bar{K}\Lambda$ and the $\bar{K}\Sigma$ invariant mass distributions. These two latter channels are ideal to study the $\Xi(1690)$  because their couplings to this resonance are much larger than that of 
$\pi\Xi$ channel, and in addition the $\Xi(1690)$ lies near these thresholds (see Tables~\ref{tab:ROB_GLN_model} and \ref{tab:Sekihara_model}) . 

\begin{figure}[tb]
\begin{center}
\includegraphics[width=8cm,bb=0 0 540 378]{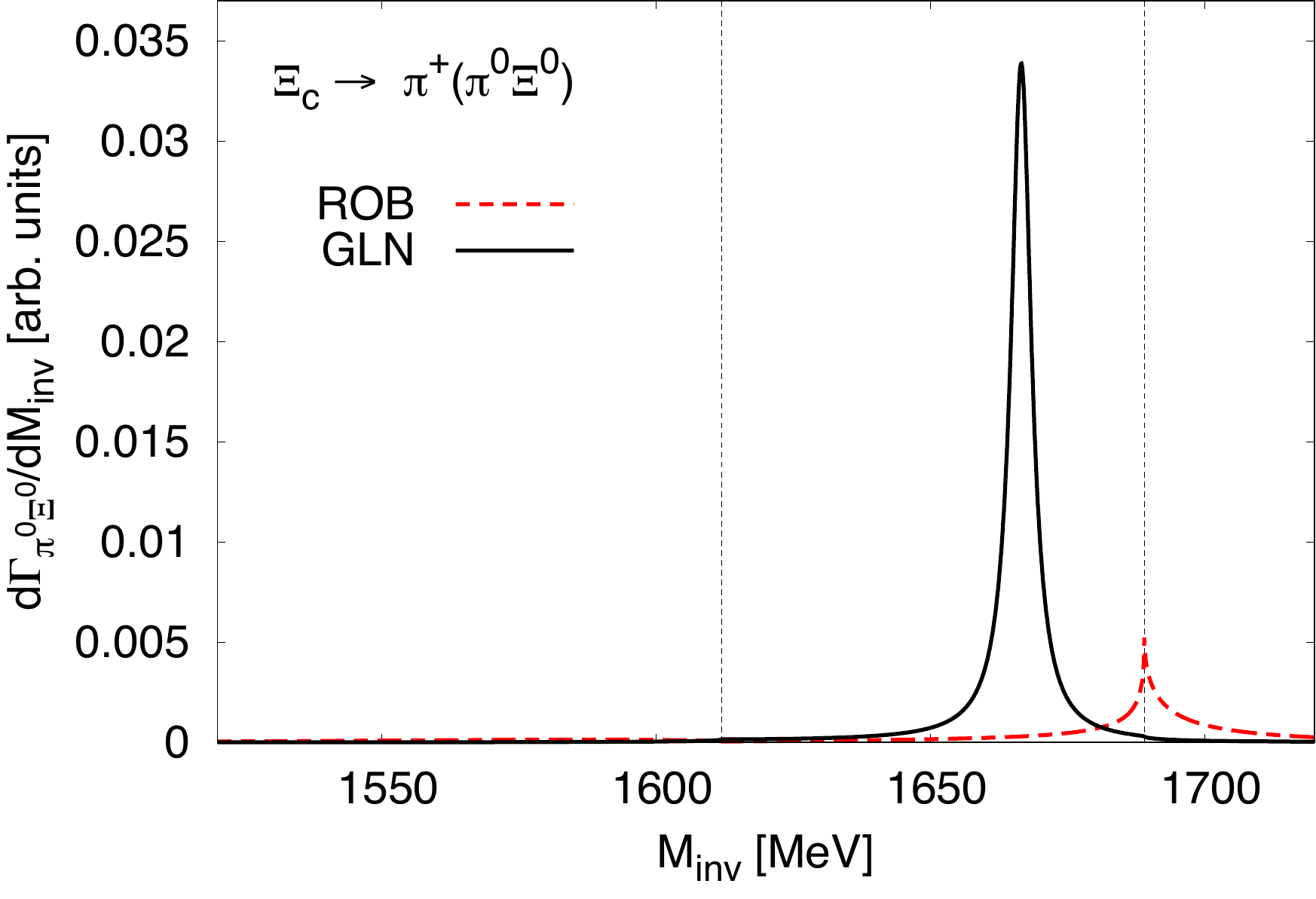}
\caption{$\pi^0\Xi^0$ invariant mass distribution obtained with the ROB (red dashed curve) and the GLN  (black solid curve) models. The vertical  lines represent the $\bar{K}\Lambda$ and $\bar{K}\Sigma$ thresholds.}
\label{fig:spe_piXi}  %\ref{fig:spe_piXi}
\end{center}
\end{figure}

In Fig.~\ref{fig:spe_piXi}, we show the $\pi^0\Xi^0$ invariant mass distribution predicted with the ROB and  GLN models. 
%%%%%%%%%%%  figure  %%%%%%%%%%%%
%%%%%%%%%%%%%%%%%%%%%%%%%%%
%%%%%%%%%%%%%%%%%%%%%%%%%%%
%%%%%%%%%%%%%%%%%%%%%%%%%%%
Though both models have the $\Xi(1620)$ resonance pole in the meson-baryon scattering amplitudes, the peak structure can be hardly seen around the $M_{\rm inv}\sim1600$ MeV region in this $\Xi_c$ decay distribution. 
%
%---- reason for no Xi(1620) peak -----
The main reason of the absence of the peak is the large decay width. Especially in the GLN model, the width is larger than 250 MeV, and such a state is difficult to see as a clear peak on the real energy axis. 
%On the other hand, the pole width is relatively smaller ($\sim 60$ MeV) in the ROB model, and the peak structure may be expected. However, in this model, another reason }
There is additional suppression of the signal related to the decay mechanism and the models for the final-state interaction. In the $\Xi_c$ decay process, the $\pi\Xi$ channel does not appear in the intermediate state as in Eqs.~\eqref{eq:MB} and \eqref{eq:MB_iso}. Hence, considering that the $\Xi(1620)$ mainly couples to the $\pi\Xi$ and $\bar{K}\Lambda$ channels (see Table~\ref{tab:ROB_GLN_model}), and  neglecting for simplicity the other channels, the decay amplitude  can be approximated in the energy region of interest for the $\Xi(1620)$  as
\begin{align}
\mathscr{M}_{\pi^0\Xi^0} \sim \frac{V_P}{3\sqrt{2}} G_{\bar{K}\Lambda}t_{\bar{K}\Lambda,\pi\Xi}. %\notag
\end{align}
The $\Xi(1620)$ appears below the $\bar{K}\Lambda$ threshold in both, ROB and GLN, chiral unitary approaches. Since there is no $\pi\Xi$ tree level contribution, the final $\pi^0\Xi^0$ state is produced only through the FSI, with its production rate just determined by the  $\bar{K}\Lambda$ loop function $G_{\bar{K}\Lambda}$.
Generally, a loop function becomes small below the threshold. Especially, in the ROB model, $G_{\bar{K}\Lambda}$ vanishes around the $\Xi(1620)$ energy region. As a consequence, it is difficult to see the $\Xi(1620)$ signal in the ROB model, in which the width is relatively small ($\sim 130$ MeV). 
%\textcolor{red}{(check the case where $\Xi(1620)$ is above the $\bar{K}\Lambda$ threshold.)}

\begin{figure}[tb]
\begin{center}
\includegraphics[width=8cm,bb=0 0 540 378]{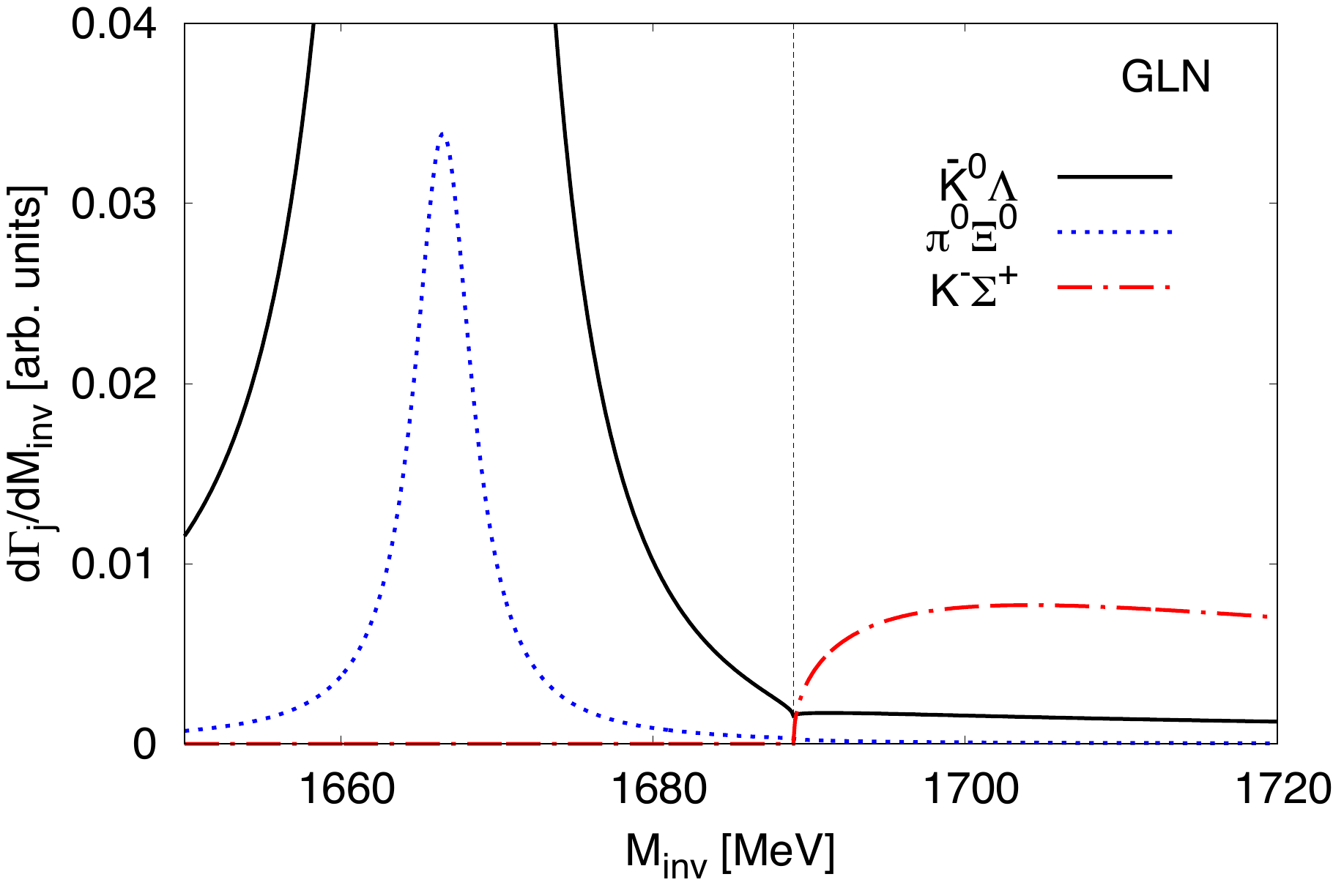}
\caption{$\bar{K}^0\Lambda$ (black solid curve),  $K^-\Sigma^+$ (red dash-dotted curve), and $\pi^0\Xi^0$ (blue dotted curve) invariant mass distributions obtained with the GLN chiral unitary approach. The vertical dashed line marks the $\bar{K}\Sigma$ threshold.}
\label{fig:spe_all}  %\ref{fig:spe_all}
\end{center}
\end{figure}

%-------- Xi(1690) peak --------------------
In sharp contrast, the $\Xi(1690)$ peak can be clearly seen in the $\pi\Xi$ distribution of the $\Xi_c$ decay, as shown in the GLN result of Fig.~\ref{fig:spe_piXi}.
%despite of the small coupling of the $\Xi(1690)$ to this channel. 
This is because in the GLN model, the decay width of the $\Xi(1690)$ is quite small and the loop functions strongly related to the $\Xi(1690)$ ($G_{\bar{K}\Sigma}$ and $G_{\eta\Xi}$) do not vanish around the $\Xi(1690)$ energy region. 
Thus, it is advisable to study  the invariant mass distributions around the $\Xi(1690)$ region for the rest of the channels. Predictions obtained with the GLN model are shown in Fig.~\ref{fig:spe_all}.
%%%%%%%%%%%  figure  %%%%%%%%%%%%
%%%%%%%%%%%%%%%%%%%%%%%%%%%
%%%%%%%%%%%%%%%%%%%%%%%%%%%
%%%%%%%%%%%%%%%%%%%%%%%%%%%
 We see the  $\Xi(1690)$ gives rise to a large peak in the $\bar{K}^0\Lambda$ channel, which  could be quite useful to extract details of this resonance.

\begin{figure}[tb]
\begin{center}
\includegraphics[width=8cm,bb=0 0 540 378]{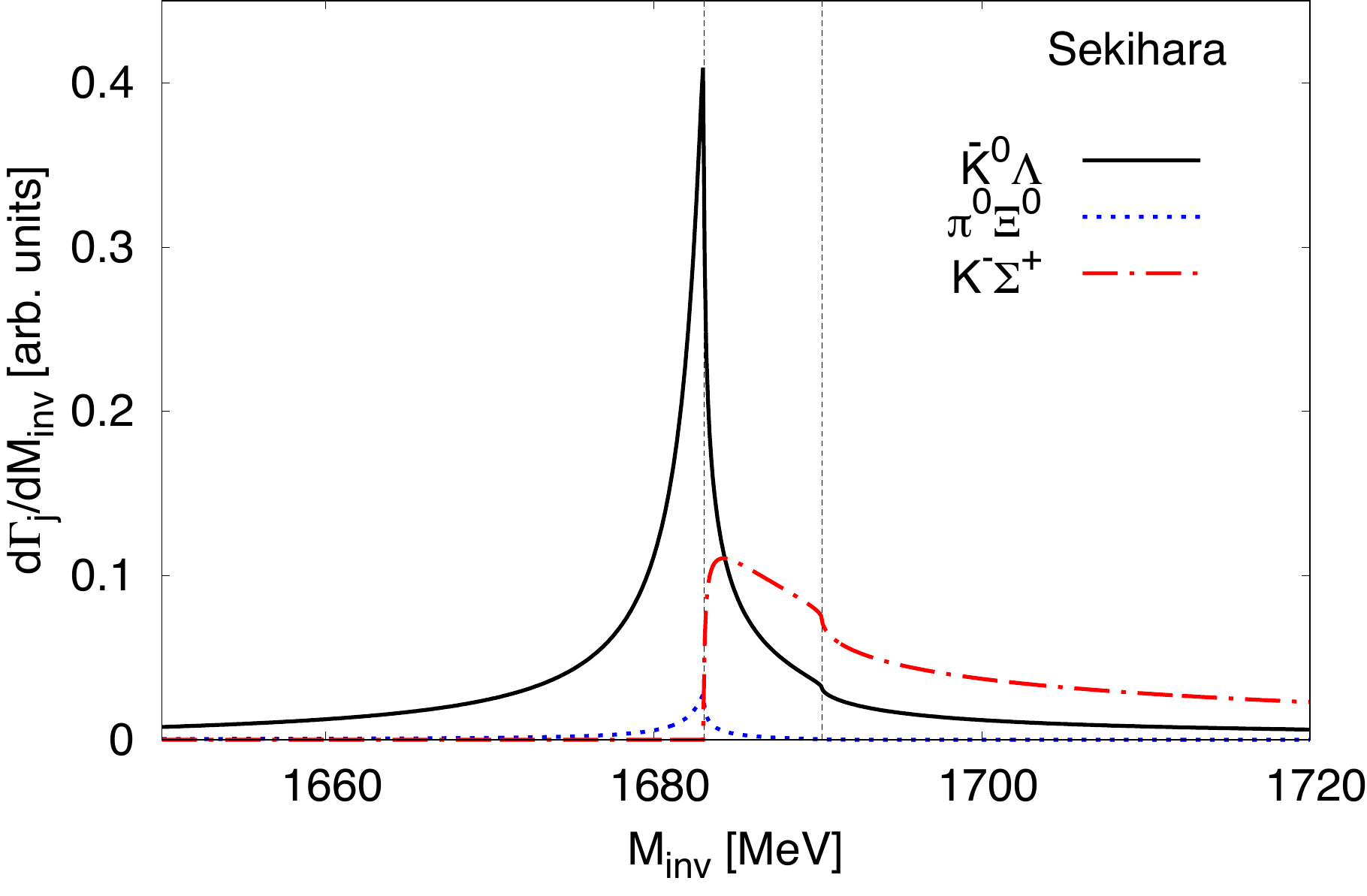}
\caption{$\bar{K}^0\Lambda$ (black solid curve),  $K^-\Sigma^+$ (red dash-dotted curve) and $\pi^0\Xi^0$ (blue dotted curve) invariant mass distributions obtained with the Sekihara model of Ref.~\cite{Sekihara:2015qqa}. The vertical dashed lines mark the $K^-\Sigma^+$ and $\bar{K}^0\Sigma^0$ thresholds.}
\label{fig:spe_seki_all}  %\ref{fig:spe_seki_all}
\end{center}
\end{figure}

For a more detailed analysis, we consider also the Sekihara model  of Ref.~\cite{Sekihara:2015qqa} to the $\Xi_c$ decay analysis. The different $MB$ invariant mass distributions are shown in Fig.~\ref{fig:spe_seki_all}. Again in this case, the $\bar{K}^0\Lambda$ distribution presents the largest $\Xi(1690)$ signal (peak). On the other hand, when comparing  with the previous distributions obtained within the GLN model, we see that the  $\Xi(1690)$ peaks in 
the $\bar{K}^0\Lambda$ and $\pi^0\Xi^0$ ($K^-\Sigma^+$) distributions predicted by the Sekihara approach are  smaller (larger) than those obtained with the GLN scheme. 
The reason is that, as mentioned above,  in the Sekihara model the $\Xi(1690)$ pole does not show up in the proper ``second Riemann sheet (SRS)'', i.e., the Riemann sheet obtained by
continuity across each of the two-body unitary cuts~\cite{Nieves:2001wt}.
%%%%%%%%%%%  figure  %%%%%%%%%%%%
%%%%%%%%%%%%%%%%%%%%%%%%%%%
%%%%%%%%%%%%%%%%%%%%%%%%%%%
%%%%%%%%%%%%%%%%%%%%%%%%%%%

Finally, we should note the existence of a cusp structure around the $\Xi(1690)$ region also in the ROB model (Fig.~\ref{fig:spe_piXi}), despite the $\Xi(1690)$ resonance is not being generated in that approach.
Indeed, the origin of this cusp is the opening of the $\bar{K}\Sigma$ threshold and in the next section, we will discuss how to distinguish this situation from a peak produced by a dynamically generated resonance.

%=====================================================================
\section{Discussion}  \label{sec:discuss}  %\ref{sec:discuss}
In the previous section, we have shown $MB$ distributions from the $\Xi_c$ decay obtained with different chiral models. Here, first we propose a method to distinguish the origin (cusp threshold effect, pole in the SRS or in a non-physical Riemann sheet) of the structures observed in the decay mass distributions. Next, we will estimate the impact of the contributions from the mechanisms  depicted in the quark-line diagrams in Fig.~\ref{fig:Xic_decay3}, which are not included in the dominant one of Fig.~\ref{fig:Xic_decay}.
Finally, we will compare  $\Xi_c^+$ and $\Xi_c^0$ decays and show that the differences among them may be useful to better understand the decay mechanisms of heavy hadrons. 

\subsection{Relation between the peak and decay ratios}
\begin{figure}[tb]
\begin{center}
\includegraphics[width=8cm,bb=0 0 540 378]{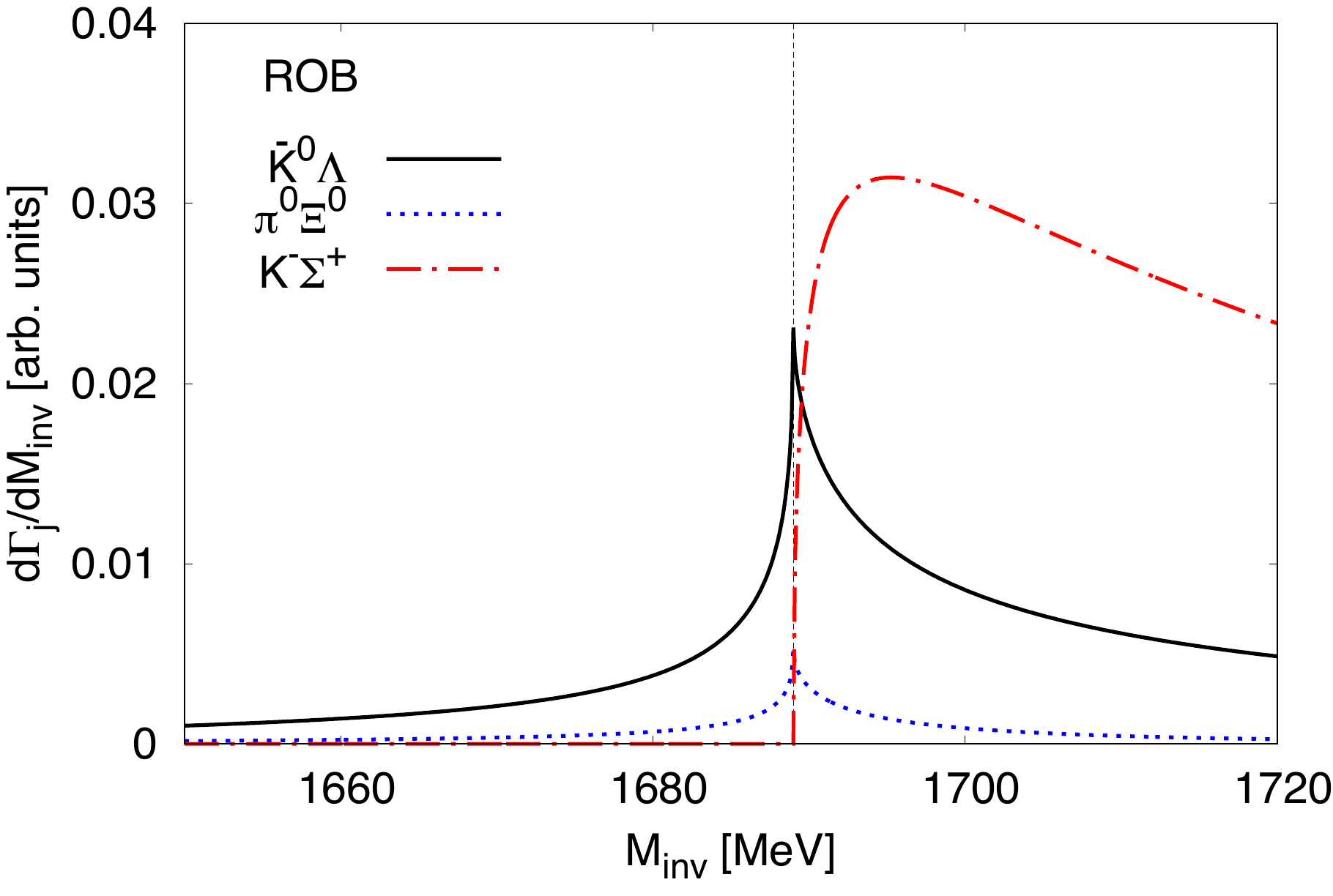}
\caption{$\bar{K}^0\Lambda$ (black solid curve),  $K^-\Sigma^+$ (red dash-dotted curve), and $\pi^0\Xi^0$ (blue dotted curve) invariant mass distributions obtained with the ROB model. The vertical dashed line indicates the $\bar{K}\Sigma$ threshold.}
\label{fig:spe_all_ROB}  %\ref{fig:spe_all_ROB}
\end{center}
\end{figure}

%-------- origin of peak ------------
As explained in Sec.~\ref{results}, there are two possibilities for the origin of the peak observed in the mass distributions around the $\Xi(1690)$ energy region. It could be produced by a pole, either in the SRS or in a nonphysical Riemann sheet,  or it might be a threshold effect. 
Here we propose that the ratios of the decay fractions around the expected position of the $\Xi(1690)$ resonance might be used to distinguish one situation from the another one. In Fig.~\ref{fig:spe_all_ROB}, we show all the invariant mass distributions with the ROB model.
%%%%%%%%%%%  figure  %%%%%%%%%%%%
%%%%%%%%%%%%%%%%%%%%%%%%%%%
%%%%%%%%%%%%%%%%%%%%%%%%%%%
%%%%%%%%%%%%%%%%%%%%%%%%%%%
Comparing these latter distributions with the those presented earlier in Figs.~\ref{fig:spe_all}, ~\ref{fig:spe_seki_all}, we see that the height of peak that appears in the $\bar{K}\Lambda$ distribution (solid curve) is much larger in the schemes with a $\Xi(1690)$ pole    (GLN and Sekihara) than in the ROB approach, where the resonance is not dynamically generated. Indeed, integrating the invariant mass distributions over the $\Xi(1690)$ region ($1650\leq M_{\rm inv}\leq1720$ MeV), we find quite different predictions for the ratios of the decay branching fractions,
\begin{align}
\frac{\Gamma_{\bar{K}^0\Lambda}}{\Gamma_{K^-\Sigma^+}} &= 15.7\ \ (\mbox{GLN}), \notag \\
\frac{\Gamma_{\bar{K}^0\Lambda}}{\Gamma_{K^-\Sigma^+}} &= 1.4\ \ (\mbox{Sekihara}), \notag \\
\frac{\Gamma_{\bar{K}^0\Lambda}}{\Gamma_{K^-\Sigma^+}} &= 0.4\ \ (\mbox{ROB}).
\end{align}
The above ratios reveal a quite large difference due to the existence or not, and in the former case to the exact nature (position) of the resonance pole. In the GLN chiral approach, the $\Xi(1690)$ is quite narrow and since the pole lies below the $K^-\Sigma^+$ threshold, the resonance does not  affect much the $K^-\Sigma^+$ channel, while its influence for the $\bar{K}^0\Lambda$ branching fraction becomes much larger.
On the other hand, in the ROB model the $\Xi(1690)$  is not generated, and the $K^-\Sigma^+$ fraction largely exceeds the $\bar{K}^0\Lambda$ one.  In the Sekihara model, because the resonance pole does not directly affect the real axis, the predicted ratio turns out to be  between those obtained in the above two cases, and $G_{\bar{K}\Lambda}$ is comparable with $G_{\bar{K}\Sigma}$.

\subsection{Contribution from other diagrams}  \label{sec:other_diagram}  %\ref{sec:other_diagram}
\begin{figure}[!h]
\begin{center}
\subfigure[]
{\includegraphics[width=8cm,bb=0 0 540 378]{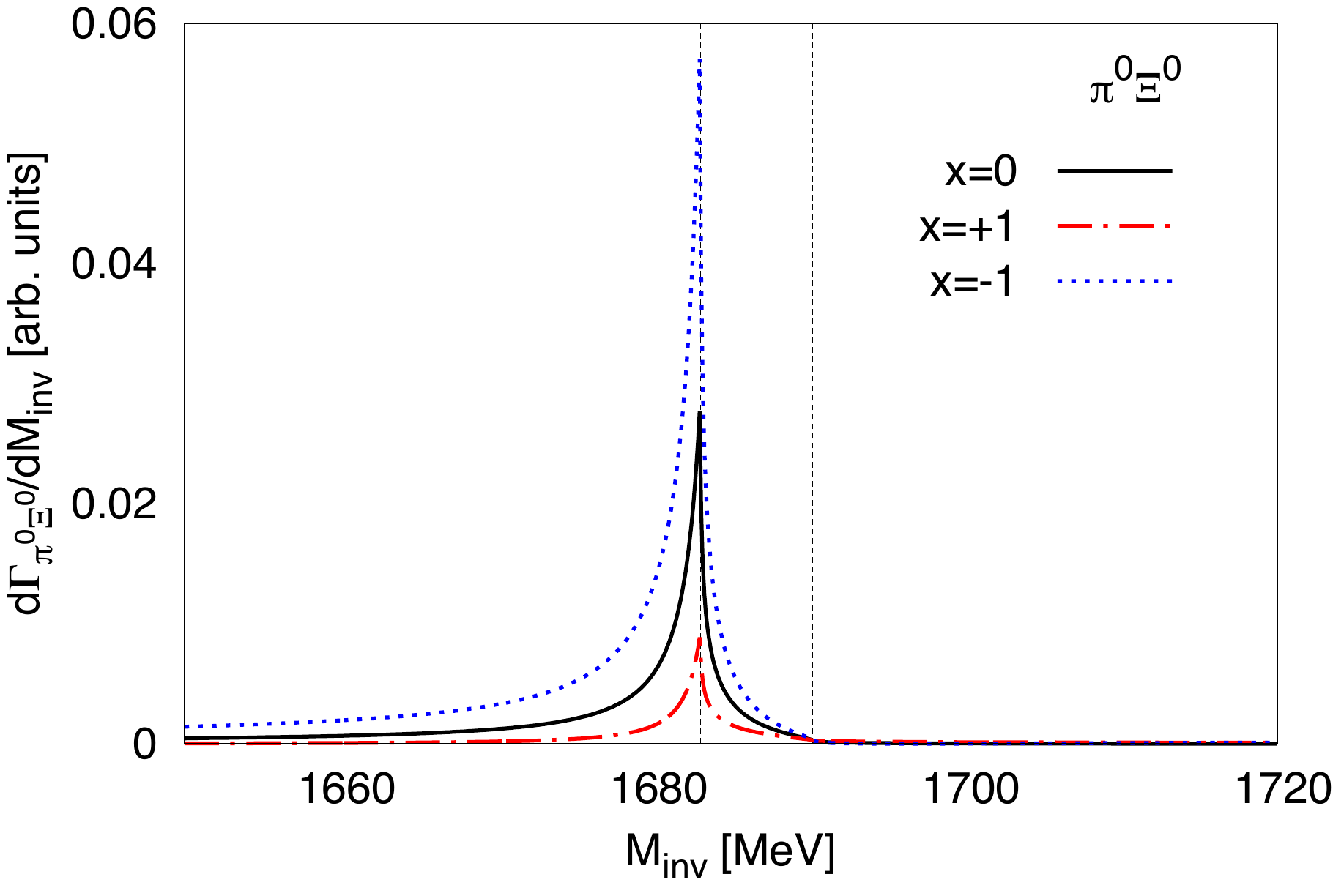}
}
\subfigure[]{
\includegraphics[width=8cm,bb=0 0 540 378]{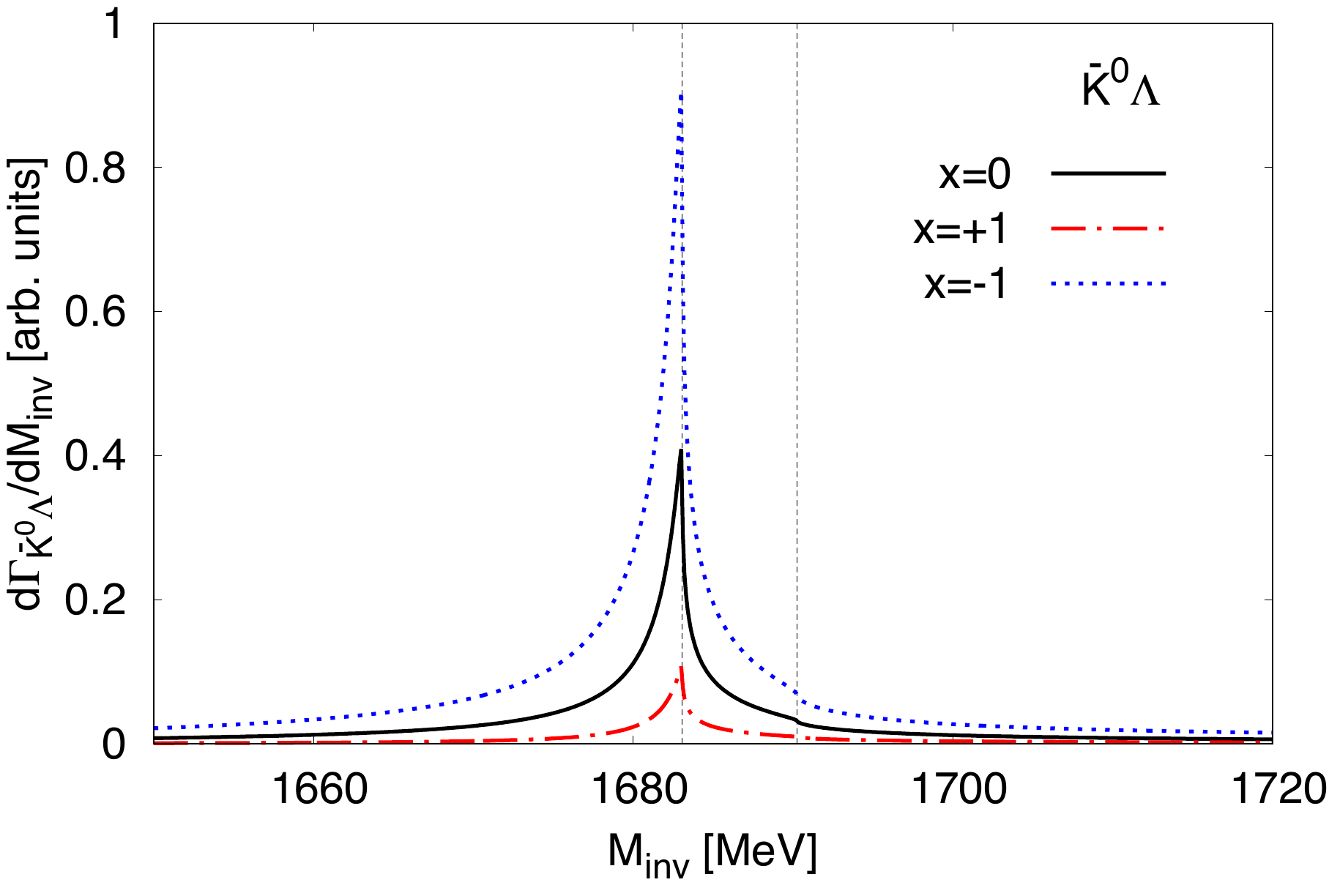}
}
\subfigure[]{
\includegraphics[width=8cm,bb=0 0 540 378]{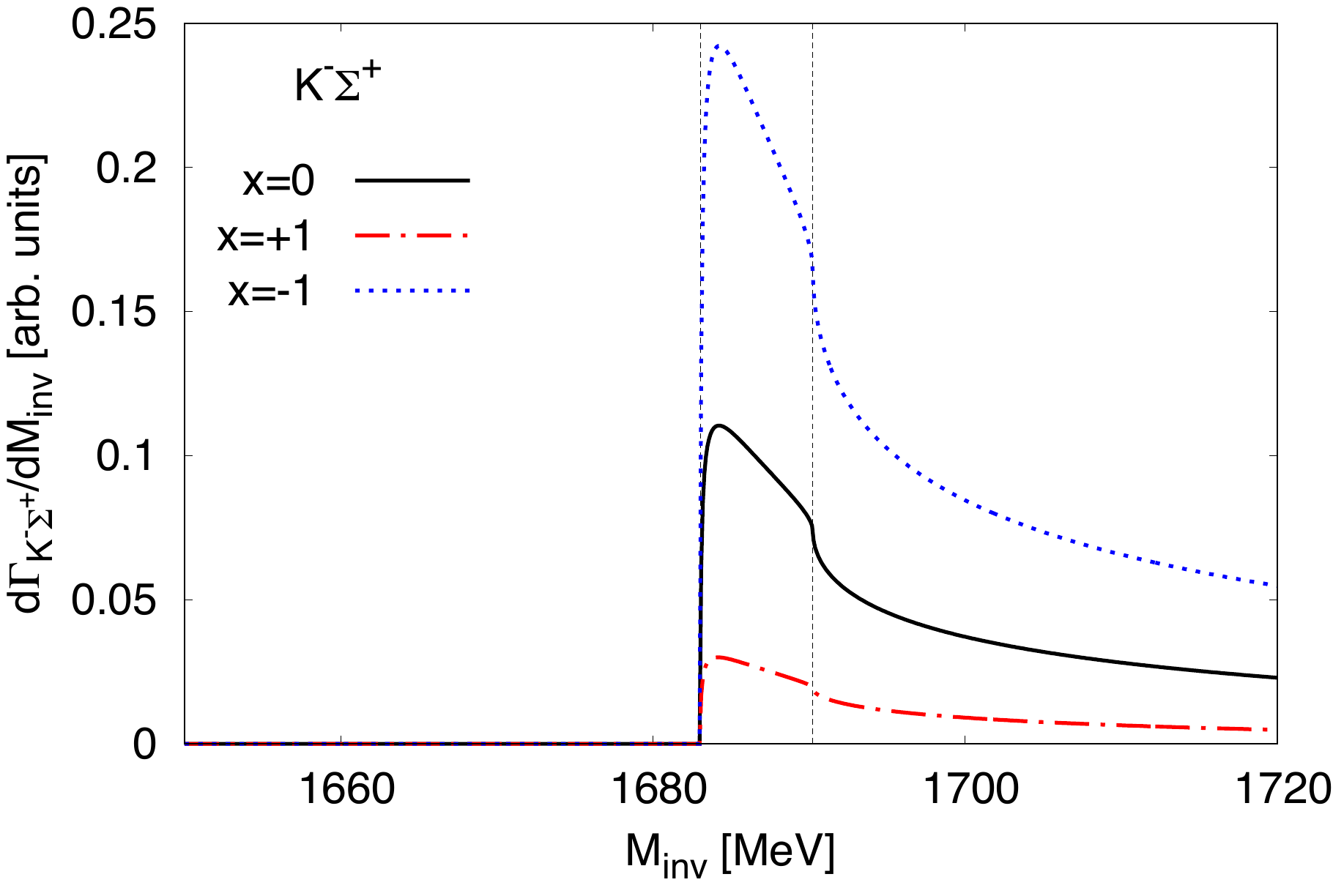}
}
\caption{(a) $\pi^0\Xi^0$, (b) $\bar{K}^0\Lambda$, and (c)  $K^-\Sigma^+$ invariant mass distributions obtained in the Sekihara model using three different values of $x$= 0 (black solid curves), $+1$ (red dash-dotted curves) and $-1$ (blue dotted curves). The weight  $x$ is defined  in Eq.~\eqref{eq:MB_add} and controls the amount of the $\eta\Xi$ component in the intermediate $MB$ state.}
\label{fig:spe_other}  %\ref{fig:spe_other}
\end{center}
\end{figure}

%------- other diagram --------
Up to now, we have set the parameter $x$ in Eq.~\eqref{eq:MB_add} to 0, which amounts to consider only the decay mechanism of the quark-line diagram depicted in Fig.~\ref{fig:Xic_decay}. Here, we try to estimate the contribution from the other mechanisms shown in Figs.~\ref{fig:Xic_decay3}(a) and \ref{fig:Xic_decay3}(b). In these latter diagrams,  the only intermediate meson-baryon state is the $\eta\Xi$, which weakly couples to  the $\Xi(1620)$ within the ROB and GLN chiral approaches (see Tables~\ref{tab:ROB_GLN_model} and \ref{tab:Sekihara_model}). Thus,  the addition of this decay mechanism should not change much the situation described above for the $\Xi(1620)$. However, the $\Xi(1690)$ has 
a substantial coupling to the $\eta\Xi$ channel, and thus the $\Xi(1690)$ production in the $\Xi_c$ decay might be affected. Predictions from the Sekihara approach for $x=-1,0$, and 1 are shown in Fig.~\ref{fig:spe_other} (mass distributions) and in  
Eq.~(\ref{eq:decayfra_other}) (decay fraction ratios, as defined above), 
%%%%%%%%%%%  figure  %%%%%%%%%%%%
%%%%%%%%%%%%%%%%%%%%%%%%%%%
\def\subfigcapskip{-3pt}
\def\subfigbottomskip{-5pt}
%%%%%%%%%%%%%%%%%%%%%%%%%%%
%%%%%%%%%%%%%%%%%%%%%%%%%%%
\begin{align}
\frac{\Gamma_{\bar{K}^0\Lambda}}{\Gamma_{K^-\Sigma^+}} &= 
\begin{cases}
1.4\ \ (x=0),  \\
1.2\ \ (x=+1),  \\
1.4\ \ (x=-1).
\end{cases}  \label{eq:decayfra_other}
%\eqref{eq:decayfra_other}
\end{align} 
Roughly speaking, the qualitative behavior and the relative weight of  the different mass distributions are not changed by the addition of the mechanisms of  Fig.~\ref{fig:Xic_decay3}, and the major differences appear in the overall height of the distributions.%
\footnote{Note that with a fine tuning of the parameter $x$, the $\Xi(1690)$ peak may accidentally disappear in the $\pi\Xi$ and $\bar{K}\Lambda$ spectra because of some  destructive interferences. }
%In the GLN and Sekihara models, such cancellation occurs for $x\sim2$ and $x\sim1$, respectively. Except for these specific values of $x$, the qualitative  discussion in the text still holds.}
We thus conclude that the results with $x=0$ presented in the previous section are reasonable, as long as the spectral shapes and the relative fractions are concerned.

\subsection{Subleading diagrams for $\Xi_c^0$ decay}   \label{sec:Xic0}  %\ref{sec:Xic0}
\begin{figure}[!h]
\begin{center}
\includegraphics[width=8cm,bb=0 0 722 411]{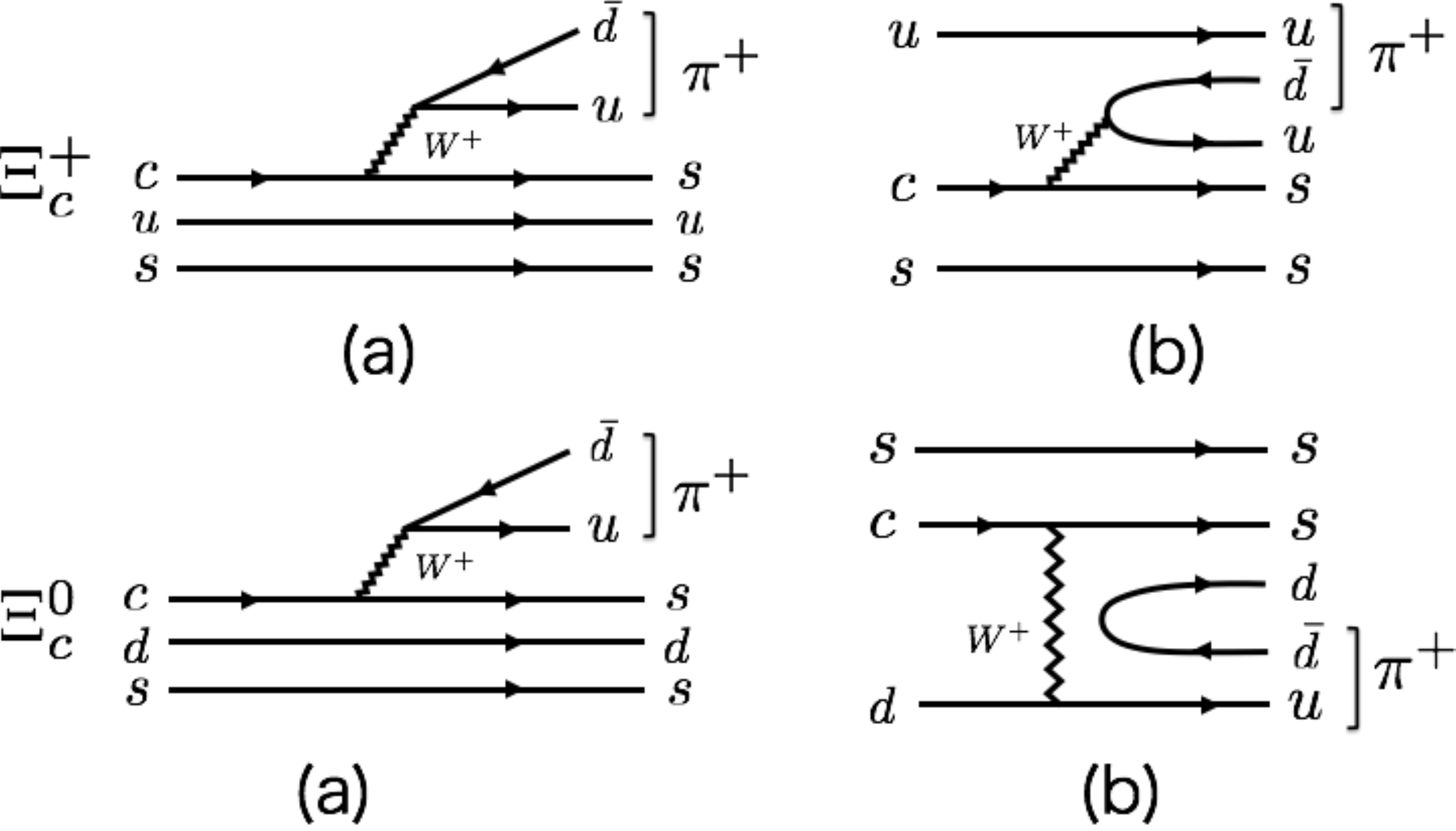}
\caption{Cabibbo favored quark-line diagrams for the $\Xi_c^+\to\pi^+(ssu)$  and $\Xi_c^0\to\pi^+(ssd)$ decays.}
\label{fig:diagram_Xic0}  %\ref{fig:diagram_Xic0}
\end{center}
\end{figure}

%-------- Xi_c^+ and Xi_c^0 ----------
Finally, in this subsection, we want to compare the $\Xi_c^0$ and $\Xi_c^+$ decays. We show in Fig.~\ref{fig:diagram_Xic0} all the Cabibbo favored quark-line diagrams, with a $\pi^+$ emitted before the $q\bar{q}$-pair insertion, for both $\Xi_c^0$  and $\Xi_c^+$ decays.  
%%%%%%%%%%%  figure  %%%%%%%%%%%%
%%%%%%%%%%%%%%%%%%%%%%%%%%%
%%%%%%%%%%%%%%%%%%%%%%%%%%%
%%%%%%%%%%%%%%%%%%%%%%%%%%%
As discussed in Sec.~\ref{sec:formulation}, diagram~(b) of the $\Xi_c^+$ decay is suppressed by color recombination factors, diquark correlations, and  kinematics when a high momentum $\pi^+$ is required. 
Similarly, the diagram~(b) of the $\Xi_c^0$ decay is also suppressed, but the topology of the subdominant diagrams is different from that of the $\Xi_c^+$ decay. 
If the resolution of the analysis is sufficient to extract the subdominant contributions, we may study the difference of the heavy hadron decay diagrams.

%=====================================================================
\section{Summary}

We have studied the $\Xi_c\to\pi^+(MB)$ decay process as a tool to study $\Xi$ resonances, such as the $\Xi(1620)$ and the $\Xi(1690)$. The $M^2_{\bar{K}\Lambda}$-$M^2_{\pi^+\bar{K}}$ Dalitz plot shows the $\Xi_c \to \pi^+(\bar{K}\Lambda)$ decay is not affected by the presence of  resonances, in sharp contrast to the $\Lambda_c\to K^+(\bar{K}\Lambda)$ decay, where the $a_0(980)$ in the $K\bar{K}$ channel considerably complicates the $\Xi$-resonance analysis. 

Taking into account Cabibbo-Kobayashi-Maskawa matrix and color suppressions, 
diquark correlations and  kinematical restrictions, we have proposed a dominant mechanism (Fig.~\ref{fig:Xic_decay}) to describe the $\Xi_c$ 
decays. This mechanism determines the relative fractions of the intermediate meson-baryon states, while the final $MB$ state interaction has been incorporated and studied using three different chiral unitary approaches. Thus, we have predicted  $MB$ invariant mass distributions in the three chiral schemes. We have seen that the $\Xi_c$ decay is not adequate to study the $\Xi(1620)$ resonance because the open channel $\pi\Xi$ is not produced a tree level, and the other possible channels are closed. However, the $\Xi(1690)$ peak can be clearly seen in the $\pi\Xi$ and $\bar{K}\Lambda$ mass distributions.

We have further analyzed the peak around the $\Xi(1690)$ energy region, because it could be produced by a pole, either in the second Riemann sheet or in a non-physical Riemann sheet,  or it might be just a threshold effect. We have shown that the ratios of the decay fractions around the expected position of the $\Xi(1690)$ resonance might be used to distinguish one situation from the another one. Comparing the several models for the final-state interaction, we have found that if the $\Xi(1690)$ pole exists below the $K^-\Sigma$ threshold, the $\bar{K}^0\Lambda$ decay fraction will largely exceed the $K^-\Sigma^+$ one, as a consequence of the quite narrow decay width of the resonance. On the other hand, if the pole is not placed in the SRS, defined by continuity with the physical sheet in  the real axis, the $\bar{K}^0\Lambda$ fraction becomes quite small and  the $K^-\Sigma^+$ mode turns out to be dominant. 

The above results are based on the mechanism shown in Fig.~\ref{fig:Xic_decay}. However, there exist other quark-line diagrams where the high energy momentum $\pi^+$ is emitted after the $\bar{q}q$ creation, which might provide also a sizable contribution. We have estimated their contribution, and found that the neglected diagrams only alter the overall height of the spectra. Thus, we have concluded that the results from the  mechanism of  Fig.~\ref{fig:Xic_decay} are reasonable, as long as only the spectral shape and the relative fractions are concerned. 

%Finally, we have compared the $\Xi_c^+$ and $\Xi_c^0$ decays, and found that the sub-dominant diagrams in these two decays are different. Future experimental data, analyzed within the scheme derived in this work, might help to clarify the relevant mechanisms of heavy hadron decays.

%=====================================================================
\section{Acknowledgments}
This work is partly supported by 
Open Partnership Joint Projects of JSPS Bilateral Joint Research Projects, JSPS KAKENHI Grant Nos. 16K17694 and 25247036, the Yukawa International Program for Quark-Hadron Sciences (YIPQS),
the National Natural Science Foundation of China under Grant Nos. 1375024 and 11522539, the Spanish Ministerio de Economia y Competitividad and European FEDER funds under contract nos.
FIS2011-28853-C02-01 and FIS2011-28853-C02-02, and the Generalitat Valenciana in the program Prometeo II-2014/068. 

%=====================================================================

%=====================================================================
\appendix
\section*{Appendix : Quark representation of hadrons}

%In this appendix, we consider the quark representation of hadrons from two points of view. Both prescriptions respect the SU(3) symmetry, and give equivalent results. 
%
%As the first prescription, we use the SU(3) pseudoscalar meson matrix as in the chiral perturbation theory~\cite{Scherer:2012xha}.
In this appendix, we consider the quark representation of hadrons, based on the SU(3) symmetry. 
For the representation of mesons, we use the SU(3) pseudoscalar meson matrix as in the chiral perturbation theory~\cite{Scherer:2012xha}. 
Respecting the SU(3) transformation, the meson degrees of freedom can be related to the quark degrees of freedom by the following equation, 
\begin{align}
M&=
\left( \begin{array}{ccc}
u\bar{u}  &  u\bar{d}  &  u\bar{s}  \\
d\bar{u}  &  d\bar{d}  &  d\bar{s}  \\
s\bar{u}  &  s\bar{d}  &  s\bar{s}  
\end{array} \right)  \notag \\
&=\left( \begin{array}{ccc}
\frac{\pi^0}{\sqrt{2}}+\frac{\eta_8}{\sqrt{6}} + \frac{\eta_1}{\sqrt{3}}  &  \pi^+  &  K^+  \\
\pi^-  &  -\frac{\pi^0}{\sqrt{2}}+\frac{\eta_8}{\sqrt{6}} + \frac{\eta_1}{\sqrt{3}} &  K^0  \\
K^-  &  \bar{K}^0  &  -\frac{2\eta_8}{\sqrt{6}} + \frac{\eta_1}{\sqrt{3}}
\end{array} \right).  \label{eq:qqbar_meson}
%\eqref{eq:qqbar_meson}
\end{align}
With regard to baryons, to obtain the similar relation, we replace the antiquarks in the meson matrix by the flavor antitriplet diquark representation suited to the mixed antisymmetric representation of the baryons which appears in Eq.~\eqref{eq:fraction_quark},
\begin{align}
\bar{u} \to \frac{1}{\sqrt{2}}(ds-sd),\ \ \bar{d} \to \frac{1}{\sqrt{2}}(su-us),\ \ \bar{s} \to \frac{1}{\sqrt{2}}(ud-du),
\end{align}
which leads to the following relation:
\begin{align}
B&=
\frac{1}{\sqrt{2}}\left( \begin{array}{ccc}
u(ds-sd)  &  u(su-us)  &  u(ud-du)  \\
d(ds-sd)  &  d(su-us)  &  d(ud-du)  \\
s(ds-sd)  &  s(su-us)  &  s(ud-du)  
\end{array} \right)  \\
&=\left( \begin{array}{ccc}
\frac{\Sigma^0}{\sqrt{2}}+\frac{\Lambda}{\sqrt{6}} + \frac{\Lambda_1}{\sqrt{3}} &  \Sigma^+  &  p  \\
\Sigma^-  &  -\frac{\Sigma^0}{\sqrt{2}}+\frac{\Lambda}{\sqrt{6}} + \frac{\Lambda_1}{\sqrt{3}}  &  n  \\
\Xi^-  &  \Xi^0  &  -\frac{2\Lambda}{\sqrt{6}} + \frac{\Lambda_1}{\sqrt{3}}
\end{array} \right).  \label{eq:qqq_baryon_pre1}
%\eqref{eq:qqq_baryon_pre1}
\end{align}
From these relations, we obtain the quark representation of the hadrons as summarized in Table~\ref{tab:MB_pre1}.%
\footnote{In order to connect the $\eta$ and $\eta'$ physical states with the $\eta_1$ and $\eta_8$ ones,  we use the standard coupling of \cite{Bramon:1992kr},
 \begin{align}
 \eta =\frac{1}{3}\eta_1 + \frac{2\sqrt{2}}{3}\eta_8,\ \ \eta'=\frac{2\sqrt{2}}{3}\eta_1 - \frac{1}{3}\eta_8.
 \end{align} }
This quark representation agrees with the one implicitly assumed in the chiral Lagrangians, as shown in Ref.~\cite{Pavao:2017cpt}.

%The second prescription is based on 
Next, we consider the quark representation from a different point of view, the assignment of the SU(3) multiplets $|N,Y,I,I_3\rangle$, where $N,\ Y,\ I,$ and $I_3$, respectively, represents the dimension of an irreducible representation, the hyper charge, the isospin, and the third component of the isospin. This is because the $MB$ fraction of heavy hadron decays can be understood as (the combination of) the Clebsch-Gordan coefficients of SU(3)~\cite{deSwart:1963pdg}.
To label the SU(3) multiplets, we use the SU(2) subgroups of SU(3), $T$ spin and $V$ spin.%
\footnote{The set of $T$ spin and $V$ spin, rather than the set of $T$ spin and $U$ spin, is chosen to be consistent with the phase convention in the hadron matrices~\cite{deSwart:1963pdg}.}
%that the phase convention of the chiral unitary approach is based on the T-spin and V-spin set~\cite{deSwart:1963pdg}.
%\textcolor{red}{to compare with the phase convention of the chiral unitary approach based on the T-spin and V-spin set~\cite{deSwart:1963pdg}} }.
The generators of these subgroups are 
\begin{align}
T_3 &=I_3=\lambda_3,\notag \\
T_+&=(\lambda_1+i\lambda_2)/2,\ \ T_-=(\lambda_1-i\lambda_2)/2,  \notag \\
V_3 &=\frac{\sqrt{3}}{4}\lambda_8+\frac{1}{4}\lambda_3, \notag \\ 
V_+&=(\lambda_4+i\lambda_5)/2,\ \ V_-=(\lambda_4-i\lambda_5)/2,
\end{align}
where $\lambda_i$ is Gell-Mann matrix. $T_{\pm}$ and $V_\pm$, respectively, correspond to the replacement,
\begin{align}
T_\pm\ :\ d\leftrightarrow u,\ \ -\bar{u}\leftrightarrow \bar{d},  \notag \\
V_\pm\ :\ s\leftrightarrow u,\ \ -\bar{u}\leftrightarrow \bar{s}. \label{eq:q_raise_low}
%\eqref{eq:q_raise_low}
\end{align}
This means that the the SU(3) multiplets $|N,Y,I,I_3\rangle$ of the quarks can be labeled as~\cite{deSwart:1963pdg}
\begin{align}
|u\rangle = |3,+\frac{1}{3},\frac{1}{2},+\frac{1}{2}\rangle&,\ \ |d\rangle = |3,+\frac{1}{3},\frac{1}{2},-\frac{1}{2}\rangle, \notag \\ 
|s\rangle =& |3,-\frac{2}{3},0,0\rangle  \notag \\
|\bar{u}\rangle = -|3^*,-\frac{1}{3},\frac{1}{2},-\frac{1}{2}\rangle&,\ \ |\bar{d}\rangle = |3^*,-\frac{1}{3},\frac{1}{2},I_z=+\frac{1}{2}\rangle, \notag \\ 
|\bar{s}\rangle =& |3^*,+\frac{2}{3},0,0\rangle.
\end{align}

%
%%%%%%  table  %%%%%%%%%%%
\begin{table*}[htb]
\begin{center}
\begin{tabular}{c|c||c|c}
Meson & $q\bar{q}$ representation & Baryon & 3$q$ representation \\ \hline
$K^+$ & $u\bar{s}$ & 
$p$ & $\frac{1}{\sqrt{2}}u(ud-du)$   \\
$K^0$ & $d\bar{s}$ & 
$n$ &$ \frac{1}{\sqrt{2}}d(ud-du)$ \\
$\pi^+$ & $u\bar{d}$ & 
$\Sigma^+$ & $\frac{1}{\sqrt{2}}u(su-us)$ \\
$\pi^0$ & $\frac{1}{\sqrt{2}}(u\bar{u}-d\bar{d})$ & 
$\Sigma^0$ & $\frac{1}{2} \left[ u(ds-sd) - d(su-us) \right]$ \\
$\pi^-$ & $d\bar{u}$ & 
$\Sigma^-$ & $\frac{1}{\sqrt{2}}d(ds-sd)$ \\
$\bar{K}^0$ & $s\bar{d}$ & 
$\Xi^0$ & $\frac{1}{\sqrt{2}}s(su-us)$ \\
$K^-$ & $s\bar{u}$ & 
$\Xi^-$ & $\frac{1}{\sqrt{2}}s(ds-sd)$ \\
%$\eta_8$ & $\frac{1}{\sqrt{6}}(u\bar{u}+d\bar{d}-2s\bar{s})$ & 
$\eta$ & $\frac{1}{\sqrt{3}}(u\bar{u}+d\bar{d}-s\bar{s})$ & 
$\Lambda$ &  $\frac{1}{2\sqrt{3}}\left[ u(ds-sd) + d(su-us) - 2s(ud-du) \right]$ \\
%$\eta_1$ & $\frac{1}{\sqrt{3}}(u\bar{u}+d\bar{d}+s\bar{s})$ & & 
$\eta'$ & $\frac{1}{\sqrt{6}}(u\bar{u}+d\bar{d}+2s\bar{s})$ & & 
\end{tabular}
\caption{Quark representation of hadrons.}
\label{tab:MB_pre1}
%\ref{tab:MB_pre1}
\end{center}
\end{table*}
%%%%%%%%%%%%%%%%%%%%%
%
%%
%%%%%%%  table  %%%%%%%%%%%
%\begin{table*}[htb]
%\begin{center}
%\begin{tabular}{c|c||c|c}
%Meson & $q\bar{q}$ representation & Baryon & 3$q$ representation \\ \hline
%$K^+$ & $u\bar{s}$ & 
%$p$ & $\frac{1}{\sqrt{2}}u(ud-du)$   \\
%$K^0$ & $d\bar{s}$ & 
%$n$ &$ \frac{1}{\sqrt{2}}d(ud-du)$ \\
%$\pi^+$ & $-u\bar{d}$ & 
%$\Sigma^+$ & $-\frac{1}{\sqrt{2}}u(su-us)$ \\
%$\pi^0$ & $-\frac{1}{\sqrt{2}}(u\bar{u}-d\bar{d})$ & 
%$\Sigma^0$ & $-\frac{1}{2} \left[ u(ds-sd) - d(su-us) \right]$ \\
%$\pi^-$ & $-d\bar{u}$ & 
%$\Sigma^-$ & $-\frac{1}{\sqrt{2}}d(ds-sd)$ \\
%$\bar{K}^0$ & $s\bar{d}$ & 
%$\Xi^0$ & $\frac{1}{\sqrt{2}}s(su-us)$ \\
%$K^-$ & $s\bar{u}$ & 
%$\Xi^-$ & $\frac{1}{\sqrt{2}}s(ds-sd)$ \\
%$\eta_8$ & $-\frac{1}{\sqrt{6}}(u\bar{u}+d\bar{d}-2s\bar{s})$ & 
%$\Lambda$ &  $-\frac{1}{2\sqrt{3}}\left[ u(ds-sd)+d(su-us)-2s(ud-du) \right]$ \\
%$\eta_1$ & $-\frac{1}{\sqrt{3}}(u\bar{u}+d\bar{d}+s\bar{s})$ & & 
%\end{tabular}
%\caption{Quark representation of hadrons from the second prescription, which respects the assignment of the SU(3)-multiplet.}
%\label{tab:MB_pre2}
%%\ref{tab:MB_pre2}
%\end{center}
%\end{table*}
%%%%%%%%%%%%%%%%%%%%%%
%%
%
%%%%%%  table  %%%%%%%%%%%
\begin{table*}[htb]
\begin{center}
\begin{tabular}{cc|c}
Meson & Baryon & $|N,Y,I,I_z\rangle$ \\ \hline
$K^+$ & $p$ 
& $|8,+1,\frac{1}{2},+\frac{1}{2}\rangle$   \\
$K^0$ & $n$ 
& $|8,+1,\frac{1}{2},-\frac{1}{2}\rangle$ \\
$\pi^+$ & $\Sigma^+$ 
& $|8,0,1,+1\rangle$ \\
$\pi^0$ & $\Sigma^0$ 
& $-|8,0,1,0\rangle$ \\
$\pi^-$ & $\Sigma^-$ 
& $-|8,0,1,-1\rangle$ \\
$\bar{K}^0$ & $\Xi^0$ 
& $|8,-1,\frac{1}{2},+\frac{1}{2}\rangle$ \\
$K^-$ & $\Xi^-$ 
& $-|8,-1,\frac{1}{2},-\frac{1}{2}\rangle$ \\
$\eta_8$ & $\Lambda$ 
& $-|8,0,0,0\rangle$ \\
$\eta_1$ & & $-|1,0,0,0\rangle$
\end{tabular}
\caption{Assignment of the SU(3)-multiplet to hadron states.}
\label{tab:assign}
%\ref{tab:assign}
\end{center}
\end{table*}
%%%%%%%%%%%%%%%%%%%%%
%
%Here, we connect quarks and hadrons. The highest weight eigenstate of the octet meson for the T and V-spin is
Here, we consider the assignment of $|N,Y,I,I_3\rangle$ to the octet quark-antiquark state.
The highest weight eigenstate of the octet meson for the $T$ and $V$ spin is
\begin{align}
|8,0,1,+1\rangle &= |3,+1/3,1/2,+1/2\rangle \otimes |3^*,-1/3,1/2,+1/2\rangle \notag \\
&= |u\bar{d}\rangle.
\end{align}
Using Eq.~\eqref{eq:q_raise_low} and the Condon-Shortley phase convention, we can determine other eigenstates.%
\footnote{
For $V^+$, we use the phase of Eqs.~(7.7) and (7.8) in Ref.~\cite{deSwart:1963pdg}.
}
For example,
\begin{align}
|8,0,1,+1\rangle&=|u\bar{d}\rangle \notag \\
\xrightarrow{V_-}& +|8,-1,1/2,+1/2\rangle=|s\bar{d}\rangle, \notag \\
|8,-1,1/2,+1/2\rangle&=s\bar{d} \notag \\
\xrightarrow{T_-} &+|8,-1,1/2,-1/2\rangle=-|s\bar{u}\rangle, \notag \\
|8,-1,1/2,-1/2\rangle&=-|s\bar{u}\rangle & \notag \\
\xrightarrow{V_+} &\frac{1}{\sqrt{2}}|8,0,1,0\rangle + \sqrt{\frac{3}{2}}|8,0,0,0\rangle = |s\bar{s}-u\bar{u}\rangle.
\end{align}
In this way, we determine the quark representation of the $|N,Y,I,I_3\rangle$ state. 
%Next, we assign meson states to $|N,Y,I,I_3\rangle$.
The SU(3) singlet state, $|1,0,0,0\rangle$, can be constructed as
\begin{align}
|1,0,0,0\rangle = -\frac{1}{\sqrt{3}} |u\bar{u}+d\bar{d}+s\bar{s}\rangle,
\end{align}
under two conditions, the orthogonality to $|8,0,1,0\rangle$ and $|8,0,0,0\rangle$ states and the Condon-Shortley phase convention, $\langle j_1,j_2+M|\otimes\langle j_2,-j_2|J,M\rangle \geq 0$,
for both the $T$ and $V$ spins.
Summarizing the assignment, we obtain
\begin{widetext}
\begin{align}
\left( \begin{array}{ccc}
u\bar{u}  &  u\bar{d}  &  u\bar{s}  \\
d\bar{u}  &  d\bar{d}  &  d\bar{s}  \\
s\bar{u}  &  s\bar{d}  &  s\bar{s}  
\end{array} \right)
&=\left( \begin{array}{ccc}
-\frac{|8,0,1,0\rangle}{\sqrt{2}}-\frac{|8,0,0,0\rangle}{\sqrt{6}} - \frac{|1,0,0,0\rangle}{\sqrt{3}}  &  |8,0,1,+1\rangle  &  |8,+1,\frac{1}{2},+\frac{1}{2}\rangle  \\
-|8,0,1,-1\rangle  &  +\frac{|8,0,1,0\rangle}{\sqrt{2}}-\frac{|8,0,0,0\rangle}{\sqrt{6}} - \frac{|1,0,0,0\rangle}{\sqrt{3}} &  |8,+1,\frac{1}{2},-\frac{1}{2}\rangle  \\
-|8,-1,\frac{1}{2},-\frac{1}{2}\rangle  &  |8,-1,\frac{1}{2},+\frac{1}{2}\rangle  &  \frac{2|8,0,0,0\rangle}{\sqrt{6}} - \frac{|1,0,0,0\rangle}{\sqrt{3}}
\end{array} \right).  \label{eq:qqbar_assign}
%\eqref{eq:qqbar_assign}
\end{align} 
\end{widetext}
Considering the chiral convention as in Eq.~\eqref{eq:qqbar_meson}, we obtain the assignment of the SU(3) multiplet to hadron states as in Table~\ref{tab:assign}. Similarly, the assignment for baryons is found from the assignment for the three-quark representation, starting from $|8,0,1,+1\rangle=u(su-us)/\sqrt{2}$, and the baryon matrix in the chiral Lagrangian as in Eq.~\eqref{eq:qqq_baryon_pre1}. The result is also shown in Table~\ref{tab:assign}.
Conversely, when the assignment in the chiral Lagrangian is taken as in Table~\ref{tab:assign}, the quark representation of hadrons is written as Table~\ref{tab:MB_pre1}.
%
%Considering the phase convention in the chiral unitary approach~\cite{Oset:1998it},
%\begin{align}
%|\pi^+\rangle&=-|8,0,1,+1\rangle,\ \ |K^-\rangle=-|8,-1,1/2,-1/2\rangle, \notag \\
%|\Sigma^+\rangle&=-|8,0,1,+1\rangle, \ \ |\Xi^-\rangle=-|8,-1,1/2,-1/2\rangle. \label{eq:ph_conv}
%%\eqref{eq:ph_conv}
%\end{align}
%we can write the mesons by the quark degrees of freedom as shown in Table~\ref{tab:MB_pre2}%
%%\footnote{
%%The $\eta_1$ is constructed to be the orthogonal state to the $\eta_8$ and $\pi^0$. Furthermore, the phase in Table~\ref{tab:MB_pre2} satisfies the the Condon-Shortley phase convention,
%%$\langle j_1,j_2+M|\otimes\langle j_2,-j_2|J,M\rangle \geq 0$,
%%for both the T and V-spin.
%%}.
%Similarly, the baryon representation can be also determined, starting from $|8,0,1,+1\rangle=u(su-us)/\sqrt{2}$, and is summarized in Tabe~\ref{tab:MB_pre2}.

%From Tables~\ref{tab:MB_pre1} and \ref{tab:MB_pre2}, it turns out that the difference of the two prescriptions is the phases of the following hadrons,
We note that at first sight, the assignment in Table~\ref{tab:assign} and the phase convention in the chiral unitary approach~\cite{Oset:1998it},
\begin{align}
|\pi^+\rangle&=-|I=1,I_z=+1\rangle,\ \ |K^-\rangle=-|I=\frac{1}{2},I_z=-\frac{1}{2}\rangle, \notag \\
|\Sigma^+\rangle&=-|I=1,I_z=+1\rangle, \ \ |\Xi^-\rangle=-|I=\frac{1}{2},I_z=-\frac{1}{2}\rangle,\label{eq:ph_conv}
%\eqref{eq:ph_conv}
\end{align}
seem to be different for the following hadrons:
\begin{align}
&\pi^+,\ \pi^0,\ \pi^-,\ \eta_8,\ \eta_1, \notag \\
&\Sigma^+,\ \Sigma^0,\ \Sigma^-,\ \Lambda.  \label{eq:diff_assign}
%\eqref{eq:diff_assign}
\end{align}
However, these assignments are physically equivalent.
The difference for the hadrons in Eq.~\eqref{eq:diff_assign} means that the phase is different for both the $s$ quark and $\bar{s}$ quark ($ud$ diquark). Because the strangeness is the conserved quantum number and the sectors with the different strangeness do not mix under the strong interaction, 
these two assignments give the same results in physical processes. 
Thus, from the usual phase convention in Eq.~\eqref{eq:ph_conv} and the assignment of the SU(3) multiplet to the quark representation as in Eq.~\eqref{eq:qqbar_assign}, we can obtain the  physically equivalent quark representation of hadrons to the one in Table~\ref{tab:MB_pre1}.
%In the present case they differ in a global, irrelevant, sign.

%=====================================================================


%\bibliographystyle{h-physrev4}
%\bibliography{refs}

\end{document}